\documentclass[11pt,english]{article}
\pdfoutput=1
\usepackage{jheppub}
\usepackage{amsfonts}
\usepackage{bbm}
\usepackage{multirow}
\usepackage{babel}
\usepackage{verbatim}
\usepackage{mathrsfs}

\newcommand{\e}{\epsilon}
\newcommand{\bL}{\bar{L}}
\renewcommand{\L}{\mathcal{L}}
\newcommand{\bcL}{\bar{\mathcal{L}}}
\newcommand{\z}{{\bar z}}
\newcommand{\h}{{\bar h}}

\newcommand{\p}{\partial}
\renewcommand{\a}{\alpha}
\renewcommand{\b}{\beta}
\renewcommand{\t}{\tau}
\newcommand{\s}{\sigma}
\newcommand{\D}{\Delta}
\newcommand{\refb}[1]{(\ref{#1})}
\newcommand{\w}{\omega}
\newcommand{\bw}{\bar{\omega}}
\newcommand{\<}{\langle}
\renewcommand{\>}{\rangle}

\newcommand{\hb}{\ensuremath{\bar{h}}}
\newcommand{\zb}{\ensuremath{\bar{z}}}
\newcommand{\K}{\ensuremath{\mathcal{K}}}
\renewcommand{\O}{\ensuremath{\mathcal{O}}}
\newcommand{\g}{\ensuremath{\mathfrak{g}}}
\newcommand{\hyp}{\ensuremath{\ _2F_1}}

\def\beaa{\begin{eqnarray*}}
\def\eeaa{\end{eqnarray*}}
\def\bee{\begin{equation*}}
\def\eee{\end{equation*}}
\def\bea{\begin{eqnarray}}
\def\eea{\end{eqnarray}}
\def\be{\begin{equation}}
\def\ee{\end{equation}}
\def\ba{\begin{align}}
\def\ea{\end{align}}
\def\tl{\tilde}
\def\vec{\overrightarrow}
\def\G{\mathcal{G}}

\title{The nuts and bolts of the BMS Bootstrap.}

\author[a]{Arjun Bagchi,} \author[b]{Mirah Gary,} 

\author[a]{and Zodinmawia.} \author{\\}

\affiliation[a]{Indian Institute of Technology Kanpur, Kalyanpur, Kanpur 208016. INDIA. \\}

\affiliation[b]{Institute for Theoretical Physics, Vienna University of Technology, A-1040 Vienna, AUSTRIA\\} 

\emailAdd{abagchi@iitk.ac.in, mgary@hep.itp.tuwien.ac.at, zodin@iitk.ac.in}

\abstract{In this paper, we elaborate on aspects of the recently introduced BMS bootstrap programme. We consider two-dimensional (2d) field theories with BMS$_3$ symmetry and extensively use highest weight representations to uncover the BMS version of crossing symmetry in 4-point functions that are constrained by symmetry. The BMS bootstrap equation is formulated and then analytic expressions for BMS blocks are constructed by looking at the limit of large central charges. These results are also applicable to 2d Galilean Conformal Field Theories through the isomorphism between the BMS$_3$ and 2d Galilean Conformal Algebras. We recover our previously obtained results in the non-relativistic limit of the corresponding ones in 2d relativistic CFTs. This provides a comprehensive check of our previous analysis. We also explore the chiral limit of BMS$_3$ where the BMS algebra reduces to a single copy of the Virasoro algebra and show that our analysis is consistent with earlier work in this direction. }

\preprint{}

\begin{document}

\maketitle

\section{Introduction}
The modern way of understanding relativistic quantum field theories (QFTs) is through renormalization group flows away from conformal field theories (CFTs). In the parameter space of all QFTs, CFTs arise as fixed points with enhanced scale and conformal symmetry. The very ambitions programme of understanding all QFTs thus is intimately related to the classification of all consistent CFTs. Conformal bootstrap \cite{Ferrara:1973yt, Polyakov:1974gs} has emerged as the leading tool in this endeavour. 

Any conformal field theory is determined by what has now come to be known as ``CFT data", viz. the spectrum of primary operators in the theory, the structure constants that are the constants of the three-point functions of primary operators not fixed by conformal invariance and the central charge of the theory (in case of 2d CFTs). But any random set of data does not constitute a consistent theory. The theory has to obey associativity of the operator algebra. Conformal bootstrap uses conformal symmetry and the consistency of the operator product expansion (OPE) to constrain possible CFTs. 

The use of the conformal bootstrap programme was initially limited to two dimensional conformal field theories \cite{Belavin:1984vu}. Here one has the additional power of infinite dimensional symmetries of the two copies of the underlying Virasoro algebra. 
\begin{subequations}\label{Vir}
\bea
&& [\L_n, \L_m]= (n-m) \L_{n+m} + \frac{c}{12} \delta_{n+m, 0} (n^3 - n) \\
&& [\bcL_n, \bcL_m]= (n-m) \bcL_{n+m} + \frac{\bar{c}}{12} \delta_{n+m, 0} (n^3 - n) \\
&& [\L_n, \bcL_m] =0
\eea
\end{subequations}
The bootstrap equation in 2d CFTs help us solve some CFTs explicitly. The analytical handle that the Virasoro symmetry provides helps put in powerful constraints on the mathematical consistency of theories in 2d. For values of the central charges between 0 and 1, there is a discrete number of CFTs with finite number of primary fields and these are called the minimal models. The bootstrap equations leads to a complete solution of the 2d minimal series. 

\medskip
Following the seminal work of Rattazzi, Rychkov, Tonni and Vichi in 2008 \cite{Rattazzi:2008pe}, who build on earlier work by Dolan and Osborn \cite{Dolan:2000ut,Dolan:2003hv}, there has been a great flurry of activity in applying conformal bootstrap techniques to spacetime dimensions higher than two. The method of conformal bootstrap has emerged as a very effective tool in calculating things like the critical exponents of Ising model or the $O(N)$ model in 3 dimensions. We refer the reader to the excellent reviews \cite{Rychkov:2016iqz, Simmons-Duffin:2016gjk} for a more detailed account of the excitement in this emerging field. See also \cite{Poland:2016chs} for a very well written overview. 

\medskip

Our present goal is to generalise the ideas and methods of the conformal bootstrap programme to theories with symmetries other than conformal invariance. In this present work, which is a continuation and elaboration of an earlier shorter piece of work \cite{Bagchi:2016geg}, we will concentrate on 2d field theories which are invariant under the following symmetry algebra:
\begin{subequations}\label{gca2d}
\bea
&& [L_n, L_m]= (n-m) L_{n+m} + c_L \delta_{n+m, 0} (n^3 - n) \\
&& [L_n, M_m]= (n-m) M_{n+m} + c_M \delta_{n+m, 0} (n^3 - n) \\
&& [M_n, M_m] =0
\eea
\end{subequations}
This algebra arises as a contraction of two copies of the Virasoro algebra \refb{Vir} and is called the 2d Galilean Conformal Algebra (GCA) \cite{Bagchi:2009my, Bagchi:2009pe}. The algebra also arises as asymptotic symmetries of 3d flat spacetimes and is called the 3d Bondi-Metzner-Sachs (BMS) algebra \cite{Bondi:1962px, Sachs:1962zza, Barnich:2006av}. This isomorphism was first noticed in \cite{Bagchi:2010zz} and goes under the name of the BMS/GCA correspondence. 

We will find that we will be able to construct, in a spirit very similar to that of CFTs, a BMS version of an OPE and then by considering four point functions, we will define the notion of BMS blocks and a BMS crossing equation. This will then lead us to the BMS bootstrap equation. In the limit of large central charge, we will find {\em{closed form}} expressions for these BMS blocks that form the basis for the solution of the bootstrap equation. We will then go on to recover all our answers as contractions of appropriate quantities in a 2d relativistic CFT. This forms a comprehensive check of our results obtained in the intrinsic method, some of which were first reported in \cite{Bagchi:2016geg}. 

To the best of our knowledge, this constitutes the first successful attempt at the construction and concrete steps towards the solution of a bootstrap equation in a theory that is not a relativistic conformal field theory. 

Our motivations for addressing field theories with the symmetry algebra \refb{gca2d} are manifold. This algebra has recently surfaced in various contexts, $viz.$ as symmetries of putative dual field theories to 3d flat space, as conformal symmetries in non-relativistic systems and also as the residual symmetry algebra on the worldsheet of the tensionless closed bosonic string \cite{Bagchi:2013bga, Bagchi:2015nca}. Below we address the first two of these applications. 

\subsection{Holography for flat spacetimes}
The notion of asymptotic symmetries is a very important concept in the study of gravitational theories, and especially in the context of holographic theories. For a fixed set of boundary conditions, the Asymptotic Symmetry Group (ASG) is the group of allowed diffeomorphisms modded out by the trivial ones (trivial diffeomorphisms are ones that lead to vanishing canonical charges). In a quantum theory of gravity, the states of the theory form representations of the ASG. The ASG also dictates the symmetries of the putative holographically dual field theory. 

Infinite dimensional ASGs turn out to be very effective in understanding aspects of the dual field theory. The most studied example of this is the ASG of AdS$_3$, which turns out to be two copies of the infinite dimensional Virasoro algebra. This leads to the conclusion that the dual field theory is a 2d CFT. The canonical analysis by Brown and Henneaux \cite{Brown:1986nw} can actually be looked upon as the birth of the AdS/CFT correspondence \cite{Maldacena:1997re}. 

Interestingly, infinite dimensional ASGs have been known to exist in the context of Minkowski spacetimes long before the discovery of Brown and Henneaux. Bondi, van der Burg, Metzner \cite{Bondi:1962px} and independently Sachs \cite{Sachs:1962zza} studied the asymptotic structure of Minkowski spacetime in 4 dimensions at its null boundary and found to their surprise that the symmetries were not dictated by the Poincare group, but an infinite dimensional group which included over and above the Poincare generators, translations of the null direction that depended on the angles of the sphere at infinity. These were called supertranslations and in spite of many efforts to do away with them, it was found that the algebra could not be truncated to just the Poincare algebra. The asymptotic symmetry algebra takes the form
\begin{subequations}\label{bms4}
\begin{align}
& [ L_n, L_m ] = (n-m) L_{m+n}, \quad [\bL_n, \bL_m] = (n-m) \bL_{n+m} \\
& [L_m, M_{r,s}] = \left( \frac{m+1}{2} - r \right) M_{m+r, s} \ , \quad [\bL_m, M_{r,s}] = \left( \frac{m+1}{2} - s\right) M_{r, m+s} \\
& [M_{r,s}, M_{t,u}] = 0
\end{align}
\end{subequations}
Here $n,m$ range from $-1$ to $+1$ while the other variables can take all integral values. The generators $M_{r,s}$ are the super-translation generators, the translations that depend on the angles of the sphere at infinity. 

Later, inspired by possible links to holography, Barnich and Troessaert \cite{Barnich:2010eb} proposed an extension of the ASG of 4d Minkowski space to include what they called super-rotations. Superrotations are group of all the conformal generators of the sphere at infinity and this extension is essentially the same as the extension of the 2d conformal algebra to include all the generators of the Virasoro algebra from the globally well-defined ones $L_{0, \pm1}$. In the above algebra, this means that $n, m$ now take all integral values. This extended ASG of 4d flatspace is now what is commonly known as the Bondi-Metzner-Sachs (BMS) group. Recently, following Strominger and collaborators \cite{Strominger:2013jfa}, a beautiful story has emerged linking BMS symmetries to soft theorems \cite{Weinberg:1965nx} and memory effects \cite{Zeldovich, Strominger:2014pwa, Pasterski:2015tva}. We refer the reader to \cite{Strominger:2017zoo} for a detailed discussion of these aspects.  

In the present paper, we are interested in the ASG of 3d Minkowski spacetimes. At null infinity, the ASG is given by the BMS$_3$ algebra \cite{Barnich:2006av}, which, as we have mentioned above, takes the form \refb{gca2d}. For Einstein gravity, the central terms are $c_L=0, \ c_M = \frac{1}{4G}$. When one considers modifications to Einstein gravity with a gravitational Chern-Simons term, a theory that goes under the name of Topological Massive Gravity, the ASG remains the same but central terms change and $c_L$ and $c_M$ are now both non-zero. Putative duals to theories with 3d gravity with asymptotically flat boundary conditions would thus be given  
by \refb{gca2d} with two non-zero central terms. A review of some progress in flat holography in general and in 3d in particular can be found in \cite{Bagchi:2016bcd, Riegler:2016hah}. An incomplete list of interesting directions that have been explored in this context are \cite{Bagchi:2012xr} -- \cite{Bagchi:2015wna}.

Our principle goal in this paper is the following. We would like to attempt to constrain 2d field theories with BMS$_3$ symmetry and hence chart out a parameter space for all possible putatively dual theories to asymptotically Minkowskian spacetimes in 3d. 

\subsection{Non-relativistic Conformal Symmetries}

We live in a world where the everyday things are governed principally by non-relativistic physics. Galilean invariant theories thus are a very good approximation for many real life applications. Thus it is vitally important to understand Galilean field theories. In analogy with relativistic QFTs, it is thus interesting to answer whether all Galilean QFTs can be understood as renormalization group flows away from fixed points governed by the analogue of conformal symmetry. Galilean Conformal Field Theories (GCFT), i.e. field theories with GCA as their symmetry algebra, arise as contractions from relativistic CFTs \cite{Bagchi:2009my}. It is thus very natural to expect that these non-relativistic fixed points in the parameter space of all Galilean QFTs will be governed by the GCA. 

Through the intriguing link of the BMS/GCA correspondence \cite{Bagchi:2010zz}, our programme thus is very useful when we consider applications to non-relativistic QFTs in 2d. This analysis, carried out to its conclusion, would thus help classify all 2d GCFTs and hence lead to an understanding of all 2d Galilean QFTs. 

It is interesting here to comment on possible extensions to higher dimensions. It has been claimed in \cite{Bagchi:2009my} that the GCA is infinite dimensional in all spacetime dimensions. This follows from the observation that the finite contracted algebra can be written in a suggestive form and given an infinite lift in any dimensions. The rather astounding claim is that the non-relativistic limit of a CFT leads to a theory which has an infinite dimensional symmetry. The infinite-dimensional GCA in any arbitrary spacetime dimensions is given by 
\begin{subequations}\label{GCA}
\bea
&& [L_n, L_m] = (n-m) L_{n+m}, \quad [M_n^i, M_m^j] =0, \\
&& [L_n, M_m^i] = (n-m) M_{n+m}^i. 
\eea
\end{subequations}
Interestingly, it has been shown that field theories like Maxwell's theory and Yang-Mills theory, which are classically conformally invariant in $D=4$, have non-relativistic versions that exhibit this infinite dimensional symmetry in the Galilean regime \cite{Bagchi:2014ysa, Bagchi:2015qcw} {\footnote{The reader is referred to \cite{Festuccia:2016caf} for a slightly different take on infinite symmetries in non-relativistic electrodynamics. Here the authors claim to have a bigger infinity of symmetries that include the GCA and in all dimensions, not only $D=4$.}}. Some recent investigations reveal that this classical symmetry enhancement is rather generic and happens in many cases where there are relativistic conformal symmetries to begin with. If there are field theories which also exhibit this infinite dimensional symmetry quantum mechanically, then these systems would be extremely interesting. They could be looked upon as closed sub-sectors in relativistic CFTs that perhaps have the promise to becoming integrable. 

In the context of the bootstrap in these higher dimensional theories, it is very possible that our methods here would generalise in a rather simple way to any dimensions. The additional power of infinite symmetries would help in the restriction of the higher dimensional theories.    

\subsection{Outline of the paper}

The rest of the paper is organised as follows. 

In Sec.~2, we give a short summary of the conformal bootstrap programme, specifically focussing on 2d CFTs. This forms a basis for the analysis we will perform for the 2d field theories with BMS symmetry. 

In Sec.~3, we look at the 2d field theories with BMS$_3$ symmetries in an intrinsic way. This means that we formulate the analogues of the conformal bootstrap analysis by relying solely on the symmetry structure of the field theory. Some of the results in this section have been reported earlier in \cite{Bagchi:2016geg}. Here we provide a detailed analysis of those results as well as some more new results which were promised but not presented in \cite{Bagchi:2016geg}. 

In Sec.~4, we first discuss the two different limits, viz. the non-relativistic and the ultra-relativistic, of the two copies of the Virasoro algebra to BMS$_3$. We then concentrate on the non-relativistic limit and recover many of the results of Sec.~3 in terms of this limit of the relativistic CFT answers. This serves as a comprehensive check of our results and also stresses the importance of the existence of this limit. 

In Sec.~5, we look at a specific subsector of the BMS$_3$ algebra, where the symmetry algebra has previously been shown to reduce to the Virasoro sub-algebra \cite{Bagchi:2009pe}. We observe that with the specific restrictions on the operator weights and central charges, the bootstrap analysis is consistent with this earlier claim. 

We conclude in Sec.~6 with a summary of the paper, some discussions and a list of future directions. 
 
\newpage

\section{The Conformal bootstrap}
In this section, we revisit some aspects of the conformal bootstrap, which we will specifically need for our analysis in the BMS bootstrap. We will confine ourselves to 2d CFTs, which are governed by two copies of the Virasoro algebra \refb{Vir}. More details can be found in the original BPZ paper \cite{Belavin:1984vu} or in some standard CFT text books \cite{DiFrancesco:1997nk, Blumenhagen:2009zz}. 

We will be work exclusively on the plane and hence the form of the generators of the 2d Virasoro algebra will be given by 
\be
\L_n = z^{n-1} \p_z, \quad \bcL_n = \bar{z}^{n-1} \p_{\bar{z}}
\ee
We define a unique vacuum state in the theory $|0\>$. One defines a state-operator correspondence in the 2d CFT:
\be
\phi(0, 0) |0\> = |\phi\>
\ee
The states in a CFT are labelled by their weights under $\L_0$ and $\bcL_0$:
\be
\L_0 |h, \h\> = h |h, \h \>, \quad \bcL_0 |h, \h \> = \h |h, \h\>
\ee
One defines a notion of primary fields as the ones which are annihilated by all positively labelled generators:
\be\label{hwV}
\L_n |h, \h\>_p = \bcL_n |h, \h\>_p =0
\ee
The representations of the Virasoro algebra, called Verma modules, are built by acting on primary fields by raising operators $\L_{-n}, \bcL_{-n}$. A general state in a CFT is given by:
\be
\phi_{p}^{\{\vec{k},\vec{\bar{k}}\}}(z, \z) = (\L_{-1})^{k_1}...(\L_{-l})^{k_l}(\bcL_{-1})^{{\bar k}_1}...(\bcL_{-j})^{{\bar k}_j}\phi_p (z,\z)\equiv \left(\L_{\vec{k}}\bcL_{\vec{\bar{k}}}\phi_p\right)(z,\z).
\ee 
 
\smallskip

\subsection{Operator product expansion}
The two and three point functions of primary states are fixed up to constants by invariance under the global part of the algebra $\L_{0, \pm1}, \bcL_{0, \pm1}$. The two-point function is given by: 
\be
\<\phi_1(z_1,\bar{z}_1)\phi_2(z_2,\bar{z}_2)\> = \frac{\frak{C}_{12}}{(z_1-z_2)^{2h}(\bar{z}_1-\bar{z}_2)^{2\bar{h}}},
\ee
The three point function of primary fields is given by: 
\be
\<\phi_1(z_1,\bar{z}_1)\phi_2(z_2,\bar{z}_2)\phi_3(z_3,\bar{z}_3)\> =  \frac{\frak{C}_{123}}{z_{12}^{-h_{123}} z_{23}^{-h_{231}} z_{13}^{-h_{312}}{\bar{z}_{12}^{-\bar{h}_{123}} \bar{z}_{23}^{-\bar{h}_{231}} \bar{z}_{13}^{-\bar{h}_{312}}}}
\ee
where $h_{ijk}=-(h_i+h_j-h_k)$. The operator product expansion (OPE) of two primary operators is given by
\be
\phi_{1}(z,\z)\phi_{2}(0,0)=\sum_{p,\{\vec{k},\vec{\bar{k}}\}}\frak{C}_{12}^{p\{\vec{k},\vec{\bar{k}}\}}z^{h_{p}-h_1-h_2+K}\,\z^{\h_{p}-\h_1-\h_2+\bar{K}}\,\phi_{p}^{\{\vec{k},\vec{\bar{k}}\}}(0,0),
\label{opeCFT}
\ee
where $K=\sum_{i}ik_{i},$ $\bar{K}=\sum_{j}j\bar{k}_{j}$. Using the OPE to find the three-point function it can be seen that $\frak{C}_{12}^{p\{0,0\}}\equiv \frak{C}_{12}^{p}=\frak{C}_{p12}$. The coefficient $\frak{C}_{12}^{p\{\vec{k},\vec{\bar{k}}\}}$ decouple as 
\be
\frak{C}_{12}^{p\{\vec{k},\vec{\bar{k}}\}}=\frak{C}_{p12}\mathcal{B}_{12}^{p\{\vec{k}\}}\bar{\mathcal{B}}_{12}^{p\{\vec{\bar{k}}\}}.
\ee
The coefficients $\mathcal{B}$ can be obtained by demanding that both sides of the OPE transform the same way under the action of $\L_m$ and $\bar{\L}_n$. These coefficients for level one and level two are shown in Table \eqref{OPE_CFT}.
\begin{table}[ht]
\def\arraystretch{1.7}
\centering
\begin{tabular}{|c|c|} 
\hline
$\mathcal{B}_{12}^{p\{1\}}=\frac{1}{2}$ & $\bar{\mathcal{B}}_{12}^{p\{\bar{1}\}}=\frac{1}{2}$
\\
\hline
$\mathcal{B}_{12}^{p\{1,1\}}=\frac{c-12h-4h_p+ch_p+8h_p^2}{4(c-10h_p+2ch_p+16h_p^2)}$ &
$\bar{\mathcal{B}}_{12}^{p\{\bar{1},\bar{1}\}}=\frac{\bar{c}-12\h-4\h_p+\bar{c}\h_p+8\h_p^2}{4(\bar{c}-10\h_p+2c\h_p+16\h_p^2)}$ \\
\hline
$\mathcal{B}_{12}^{p\{2\}}=\frac{2h-h_p+4hh_p+h_p^2}{c-10h_p+2ch_p+16h_p^2}$ &
$\bar{\mathcal{B}}_{12}^{p\{\bar{2}\}}=\frac{2\h-\h_p+4\h\h_p+\h_p^2}{\bar{c}-10\h_p+2\bar{c}\h_p+16\h_p^2}$\\
\hline
\end{tabular}
\caption{Coefficients of OPE at level 1 and level 2.}
\label{OPE_CFT}
\end{table}

\subsection{Conformal blocks and crossing symmetry}
Invariance under global conformal symmetry is not enough to fix the four-point functions of primary fields. Global invariance can help fix the form of these correlators up to a function of the conformally invariant cross ratios given below. The four-point function has the form
\bea
\<\prod_{i=1}^{4}\phi_{i}(z_{i},\z_{i})\>&=&\prod_{1\leq i<j\leq4}z_{ij}^{\sum_{k=1}^{4}h_{ijk}/3}\bar{z}_{ij}^{\sum_{k=1}^{4}\h_{ijk}/3}\mathcal{F}_{CFT}(z,\bar{z}),
\eea
where $\mathcal{F}_{CFT}(z,\bar{z})$ is an arbitrary coefficients of the cross-ratios $z$ and $\bar{z}$
\be
z=\frac{(z_1-z_2)(z_3-z_4)}{(z_1-z_3)(z_2-z_4)},\,\,\,\z=\frac{(\z_1-\z_2)(\z_3-\z_4)}{(\z_1-\z_3)(\z_2-\z_4)}.  
\ee
We can always do a global conformal transformation such that 
\be 
\{(z_i,\z_i)\}\rightarrow \{(\infty,\infty),(1,1),(z,\z),(0,0)\}.
\ee 
So we define
\be
\lim_{z_1,\z_1\rightarrow \infty}z_1^{2h_1}\z_1^{2\bar{h}_1}\<\phi_1(z_1,\z_1)\phi_2(1,1)\phi_3(z,\z)\phi_4(0,0)\>=\mathcal{G}_{34}^{21}(z,\z), 
\ee
where
\be
 \mathcal{G}_{34}^{21}(z,\z)=\<h_1,\h_1|\phi_2(1,1)\phi_3(z,\z)|h_4,\h_4\>.
\ee
Using the OPE on $\phi_3$ and $\phi_4$ inside the correlator, the function $ \mathcal{G}_{34}^{21}(z,\z)$ can be written in terms of three-point functions of primaries and their descendants. Specifically, we have
\be
 \mathcal{G}_{34}^{21}(z,\z)=\sum_p \frak{C}_{34}^p \frak{C}_{12}^p \mathcal{A}_{34}^{12}(p|z,\z). 
\ee
The blocks $\mathcal{A}_{34}^{21}(p|z,\z)$ factorizes into a holomorphic and an anti-holomorphic parts
\be
\mathcal{A}_{34}^{21}(p|z,\z)=\mathcal{F}_{34}^{12}(p|z)\bar{\mathcal{F}}_{34}^{21}(p|\z), 
\ee 
where
\be
\mathcal{F}_{34}^{21}(p|z,\z)=z^{h_p-h_3-h_4}\sum_{\{k\}}\mathcal{B}_{34}^{p\{k\}}z^K\frac{\<h_1|\phi_2(1)\L_{-k_1}...\L_{-k_N}|h_p\>}{\<h_1|\phi_2(1)|h_p\>}. 
\ee

Inside the correlator we can move the operators around which does not matter except for fermions which would introduce a sign. So apart from  $ G_{34}^{21}(z,\z)$ we may also define
\be
\lim_{z_1,\z_1\rightarrow \infty}z_1^{2h_1}\z_1^{2\bar{h}_1}\<\phi_1(z_1,\z_1)\phi_4(1,1)\phi_3(z,\z)\phi_2(0,0)\> = \mathcal{G}_{32}^{41}(z,\z)=\<h_1,\h_1|\phi_4(1,1)\phi_3(z,\z)|h_2,\h_2\>.
\ee
It can be seen from their definition that
\be\label{confcross}
\mathcal{G}_{34}^{21}(z,\z)=\mathcal{G}_{32}^{41}(1-z,1-\z).
\ee
If we expand both sides in term of the conformal blocks we have the bootstrap equation
\be
\sum_p \frak{C}_{34}^p \frak{C}_{21}^p \mathcal{A}_{34}^{21}(p|z,\z)=\sum_q \frak{C}_{41}^q \frak{C}_{32}^q \mathcal{A}_{32}^{41}(p|1-z,1-\z). 
\ee

\subsection{Global Conformal blocks}
The large central charge limit of the Virasoro algebra simplifies a lot of the analysis in 2d CFTs and has been recently pursued actively starting from \cite{Hartman:2013mia}. There are particular simplifications for the conformal blocks. 
The global conformal block is the large central charge limit of the Virasoro block \cite{Fitzpatrick:2014vua,Fitzpatrick:2015zha}. This is given by
\bea
\frak{g}_{34}^{21}(p|z,\z)&=&z^{h_p-h_3-h_4}\sum_{k}\mathcal{B}_{34}^{p\{k\}}z^k\frac{\<h_1|\phi_2(1)\L_{-1}^k|h_p\>}{\<h_1|\phi_2(1)|h_p\>}\cr
&&\times \z^{\h_p-\h_3-\h_4}\sum_{\bar{k}}\bar{\mathcal{B}}_{34}^{p\{\bar{k}\}}\z^{\bar{k}}\frac{\<\h_1|\phi_2(1)\bar{\L}_{-1}^{\bar{k}}|\h_p\>}{\<\h_1|\phi_2(1)|\h_p\>}.
\eea
The closed form expression of this can be obtained by using the constraint that both sides of the OPE \eqref{opeCFT} transform the same way under the quadratic Casimirs \cite{Dolan:2000ut,Dolan:2003hv}
\be
\mathcal{C}=\L_0^2-\frac{1}{2}(\L_1\L_{-1}+\L_{-1}\L_1),\,\,\,\bar{\mathcal{C}}=\bar{\L}_0^2-\frac{1}{2}(\bar{\L}_1\bar{\L}_{-1}+\bar{\L}_{-1}\bar{\L}_1),
\ee
of the global subgroup generated by $\{\L_{0,\pm 1},\bar{\L}_{0,\pm 1}\}$.
For simplicity, take all the external operators to be identical and to be a scalar with dimension $\Delta_\phi$. We may write the global block as
\be
\frak{g}_{\Delta_{\phi}}(p|z,\z)=z^{-\Delta_{\phi}} \bar{z}^{-\Delta_{\phi}}\mathcal{K}_{h_p,\bar{h_p}}(z,\bar{z}).
\ee
The constraint that both sides of the OPE transform the same way under the two quadratic Casimirs gives two differential equations for $\mathcal{K}_{h_p,\bar{h_p}}(z,\bar{z})$ 
\bea
&&\left[z^2(1-z)\p_z^2-z^2 \p_z\right]\mathcal{K}_{h_p,\bar{h}_p}(z,\bar{z})=h_p(h_p+1)\mathcal{K}_{h_p,\bar{h}_p}(z,\bar{z}),\cr 
&&\left[\bar z^2 (1-\bar z)\p_{\bar z}^2-\bar z^2 \p_{\bar z}\right]\mathcal{K}_{h_p,\bar{h}_p}(z,\bar{z})=\bar{h}_p(\bar{h}_p+1)\mathcal{K}_{h_p,\bar{h}_p}(z,\bar{z}).
\label{diff_eqn_cft}
\eea
Assuming $\mathcal{K}$ holomorphically factorizes
\begin{equation}
  \K_{h_p,\hb_p}(z,\zb) = \K_{h_p}(z)\bar{\K}_{\hb_p}(\zb),
\end{equation}
the solution are given in terms of gauss hypergeometric function 
\begin{equation}
  \K_{h_p}(z) = \alpha z^{-h_p} \hyp\left(-h_p,-h_p;-2h_p; z\right) + \gamma z^{h_p+1} \hyp\left(h_p+1,h_p+1;2h_p+2;z\right)
\end{equation}
and similarly for the anti-holomorphic sector.
We expand around $z=0$ and match to the boundary conditions $\beta_{\phi\phi}^{p,0}=1$, $\beta_{\phi\phi}^{p,1}=\frac{1}{2}$,
\begin{equation}
  \K_{h_p}(z) = \alpha z^{-h_p}\left(1 - \frac{h_p}{2}z + \O(z^2)\right) + \gamma z^{h_p+1}\left(1+\frac{h_p+1}{2}z + \O(z^2)\right). 
\end{equation}
The final result for the global block is \cite{Dolan:2000ut,Dolan:2003hv}
\begin{equation}
\g_{\Delta_\phi}(p|z,\zb) = z^{h_p-2h_\phi}\zb^{\hb_p-2\hb_\phi}\hyp\left(h_p,h_p;2h_p;z\right)\hyp\left(\hb_p,\hb_p;2\hb_p;\zb\right).
\end{equation}
More generally,
\begin{equation}
\g^{21}_{34}(p|z,\zb) = z^{h_p-h_1-h_2}\zb^{\hb_p-\hb_1-\hb_2}\hyp(h_p+h_{12},h_p+h_{34};2h_p; z)\hyp(\hb_p+\hb_{12},\hb_p+\hb_{34};2\hb_p;\zb),
\label{CFF_global_blocks}
\end{equation}
where $h_{ij}=h_i-h_j,\,\h_{ij}=\h_i-\h_j$. 

\newpage

\section{Bootstrapping BMS symmetries: intrinsic analysis}
In this section, we will construct the bootstrap programme for field theories with BMS symmetries through an intrinsic method. This just means that we will be inspired by the methods of 2d CFTs that we outlined in the previous section, but there will be many crucial differences, as the symmetry algebra \refb{gca2d} is fundamentally different from two copies of the Virasoro algebra \refb{Vir}. In the subsequent section we will provide a limiting analysis where we consider the contraction of \refb{Vir} to \refb{gca2d} and we will recover some of the answers of this section through the limit. Some of the central results of this section have already appeared in \cite{Bagchi:2016geg}. In this paper, and particularly in this section, we provide a much more detailed exposition of the basic analysis presented earlier. There are a number of new mathematical details and results that are presented here. 

\subsection{Highest weight representations}
We consider 2d field theories that are invariant under the BMS$_3$ algebra. We will call the directions of the field theory $(u, v)$. We will be interested in representation of the algebra \refb{gca2d} given by
\be\label{planevec}
L_n = - u^{n+1} \p_u - (n+1) u^n v \p_v, \quad M_n = u^{n+1} \p_v
\ee
This will be called the ``plane'' representation of the BMS$_3$ algebra. 

The states of the BMS invariant 2d field theory are by their weights under $L_0$.  Since $M_0$ and $L_0$ commute, the states get an additional label under $M_0$ as well. 
\be
L_0 |\D, \xi \> = \D |\D, \xi \>, \quad M_0 |\D, \xi \> = \xi |\D, \xi \>
\ee
Like in usual 2d CFTs, we will build the representation theory by first defining BMS primary operators. We do this by demanding that the spectrum (defined with respect to $\D$) be bounded from below. Then the BMS primary operators $|\D, \xi \>_p$ are the ones for which 
\be\label{hwB}
L_n |\D, \xi \>_p = M_n |\D, \xi \>_p= 0 \quad \forall n >0. 
\ee
We will assume a state-operator correspondence in the case of BMS field theories as well. While this is not strictly necessary for our analysis, it would be good to have the freedom to talk about operators and states interchangeably.  The BMS modules, very much like the Verma modules in the case of the Virasoro algebra, are built by acting creation operators on the BMS primary states. 

\subsection{Operator product expansion}
The main objects of physical interest in field theory are the correlation functions. If we know all the correlation functions, we may say that we have completely solved the theory. In finding the form of these functions, symmetries play an important role. It is interesting to know which part of the correlation functions is fixed by symmeties alone and what other parts depend on the dynamics of the theory.  In particular, for BMS-invariant theories, the co-ordinate dependence of the two-point and three-point functions are completely fixed simply by invariance under the global subgroup of the BMS group i.e., co-ordinate transformation generated by $L_{0,\pm1},M_{0,\pm1}$. The two-point function is given by \cite{Bagchi:2009ca,Bagchi:2009pe}
\bea
\<\phi_1(u_1,v_1)\phi_2(u_2,v_2)\> = \delta_{12} \ u_{12}^{-\Delta} e^{\frac{ 2 \xi_1 v_{12}}{u_{12}}}\delta_{\Delta_1\Delta_2}\delta_{\xi_1\xi_2}
\eea
The normalisation of the 2-point function has been fixed to $\delta_{12}$. The three-point function is given by
\bea
\<\phi_1(u_1,v_1)\phi_2(u_2,v_2)\phi_3(u_3,v_3)\> = C_{123} \ u_{12}^{\D_{123}} u_{23}^{\D_{231}} u_{13}^{\D_{312}}e^{-\xi_{123}\frac{v_{12}}{u_{12}}}e^{-\xi_{231}\frac{v_{23}}{u_{23}}} e^{-\xi_{312}\frac{v_{13}}{u_{13}}}.
\label{eqn:3-pt}
\eea
Here $\D_{ijk} = - (\D_i + \D_j - \D_k)$ and $\xi_{ijk}$ is defined similarly. $C_{123}$ is an arbitrary parameter called the structure constant. It is not fixed by symmetry but depends on the dynamics (or the details) of the field theory under consideration. So, if these constants are given to us, we can completely determine the three-point function by symmetry consideration alone. 

We can also consider higher correlation functions and see how much of their form are fixed by symmetry alone and what other dynamical inputs are needed to fixed the rest.  Now, all information about the correlation functions are contained in the operator product algebra, which gives the operator product expansion (OPE) of two primary fields as summation over the primaries and towers of their descendants. So, in order to know how the correlation functions are constrained by symmetries, it is enough to study constraints on the OPE. Indeed, considering these symmetries, we make the following ansatz for the OPE of two primary fields with weights $(\D_1,\xi_1)$ and $(\D_2,\xi_2)$
\bea
&&\phi_{1}(u_1,v_1)\phi_{2}(u_2,v_2)=\cr
&& \quad \sum_{p,\{\vec{k},\vec{q}\}} u_{12}^{-\D_{1}-\D_{2}+\D_{p}}\,e^{(\xi_{1}+\xi_{2}-\xi_{p})\frac{v_{12}}{u_{12}}}\left(\sum_{\a=0}^{K+Q} C_{12}^{p\{\vec{k},\vec{q}\},\a}u_{12}^{K+Q-\a}v_{12}^{\a}\right)\,\phi_{p}^{\{\vec{k},\vec{q}\}}(u_2,v_2).\cr
&&
\label{ope1}
\eea
Our notation is that for vectors $\vec{k}=(k_1,k_2,...k_r)$ and $\vec{q}=(q_1,q_2,...q_s)$, 
descendant fields $\phi_{p}^{\{\vec{k},\vec{q}\}}(u_2,v_2)$ are given by
\bea
\phi_{p}^{\{\vec{k},\vec{q}\}}(u,v)&=&\left((L_{-1})^{k_1}...(L_{-l})^{k_l}(M_{-1})^{q_1}...(M_{-j})^{q_j}\phi_p\right)(u,v)\cr
&\equiv &\left(L_{\vec{k}}M_{\vec{q}}\phi_p\right)(u,v),
\eea
where $K=\sum_{l}lk_{l},\,Q=\sum_{j}jq_{j}$. So, $\phi_{p}^{\{\vec{k},\vec{q}\}}(u,v)$ is a descendant field at level $K+Q$. For ease of calculation we can take the point $(u_2,v_2)$ in \eqref{ope1} to be the origin, giving us
\bea
\phi_{1}(u,v)\phi_{2}(0,0)&=&\sum_{p,\{\vec{k},\vec{q}\}} u^{-\D_{1}-\D_{2}+\D_{p}}\,e^{(\xi_{1}+\xi_{2}-\xi_{p})\frac{v}{u}}\left(\sum_{\a=0}^{K+Q} C_{12}^{p\{\vec{k},\vec{q}\},\a}u^{K+Q-\a}v^{\a}\right)\,\phi_{p}^{\{\vec{k},\vec{q}\}}(0,0)\cr
&\equiv & \rm{LHS}
\label{ope}
\eea
Here the form of the factor $u^{-\D_{1}-\D_{2}+\D_{p}}\,e^{(\xi_{1}+\xi_{2}-\xi_{p})\frac{v}{u}}$
is fixed by the requirement that the OPE gives the correct two-point
function and the factor $\sum_{\a=0}^{K+Q} C_{12}^{p\{\vec{k},\vec{q}\},\a}u^{K+Q-\a}v^{\a}$ is to ensure that both sides of the OPE transform
the same way under the action of $L_{0}$. To verify this second requirement, let us act
both sides of \eqref{ope} on the the vacuum $|0,0\>$ and then see the action of
$L_{0}$ on the resulting state. On the LHS we have,
\bea
L_{0}\phi_{1}(u,v)\phi_{2}(0,0)|0,0\> & = & ([L_{0},\phi_{1}(u,v)]+\phi_{1}(u,v)L_{0})\phi_{2}(0,0)|0,0\> \cr
& = & (u\p_{u}+v\p_{v}+\D_{1}+\D_{2})\phi_{1}(u,v)\phi_{2}(0,0)|0,0\>.
\eea
So, if the OPE is correct, the RHS of equation \eqref{ope} must also transform as above
\bea
L_{0}({\rm{RHS}})=(u\p_{u}+v\p_{v}+\D_{1}+\D_{2}){\rm{RHS}} 
 \label{L0RHS}
\eea
If we use the commutator
$L_{0}L_{\vec{k}}M_{\vec{q}}=L_{\vec{k}}M_{\vec{q}}L_{0}+(K+Q)L_{\vec{k}}M_{\vec{q}}$ on the LHS of the above equation we have
\bea
&&L_{0}(\rm{RHS})=\cr
&&\quad \sum_{p,\{\vec{k},\vec{q}\}} u^{-\D_{1}-\D_{2}+\D_{p}}\,e^{(\xi_{1}+\xi_{2}-\xi_{p})\frac{v}{u}}\left(\sum_{\a=0}^{K+Q}C_{12}^{p\{\vec{k},\vec{q}\},\a}u^{K+Q-\a}v^{\a}\right)(\D_{p}+K+Q)L_{\vec{k}}M_{\vec{q}}|\D_{p},\xi_{p}\>. \cr
 &&
\eea
It can also be easily checked that 
\bea
&&(u\p_{u}+v\p_{v}+\D_{1}+\D_{2}){\rm{RHS}}=\cr
&&\quad \sum_{p,\{\vec{k},\vec{q}\}} u^{-\D_{1}-\D_{2}+\D_{p}}\,e^{(\xi_{1}+\xi_{2}-\xi_{p})\frac{v}{u}}\left(\sum_{\a=0}^{K+Q}C_{12}^{p\{\vec{k},\vec{q}\},\a}u^{K+Q-\a}v^{\a}\right)(\D_{p}+K+Q)L_{\vec{k}}M_{\vec{q}}|\D_{p},\xi_{p}\>.\cr
 &&
\eea
Thus, equation \eqref{L0RHS} is satisfied, which means that both sides of OPE transform the same way under the action
of $L_{0}$. Furthermore, using the OPE inside the three-point functions and comparing the
coefficients with \eqref{eqn:3-pt} it can be seen that
\be
C_{12}^{p\{0,0\},0}\equiv C_{12}^{p}=C_{p12}.
\ee
Therefore, we will rewrite $C_{12}^{p\{\vec{k},\vec{q}\},\a}$ as 
\be
C_{12}^{p\{\vec{k},\vec{q}\},\a}=C_{12}^{p}\b_{12}^{p\{\vec{k},\vec{q}\},\a},
\ee
where, by convention,
\be
\b_{12}^{p\{0,0\},0}=1.
\ee

The coefficients $\b_{12}^{p\{\vec{k},\vec{q}\},\a}$ can be calculated by demanding that both sides of \eqref{ope} transform the same way under the other generators $L_m$ and $M_n$. Thus, the form of the OPE is completely constrained by symmetries to depend only on external inputs, such as the structure constants, the spectrum of primary operators appearing in the OPE, and the central charge. In other words, if these dynamical inputs are given to us, we can use symmetries to calculate all the correlation functions in a BMS-invariant field theory. These dynamical inputs can be used to classify and completely specify a given BMS-invariant field theory. However, any random sets of these dynamical inputs need not constitute a consistent field theory; they must satisfy a constrain equation given by the BMS bootstrap equation, which arises as a condition for the associativity of the operator product algebra. 

\subsection{Recursion relations}
Now let us try to find recursion relations for evaluating the coefficients $\b_{12}^{p\{\vec{k},\vec{q}\},\a}$. For the sake of simplicity we will consider the case $\D_{1}=\D_{2}=\D,\,\xi_{1}=\xi_{2}=\xi$.
Applying both sides of equation \eqref{ope} to the vacuum we have 
\be
\phi_{1}(u,v)|\D,\xi\>=\sum_{p}u^{-2\D+\D_{p}}\,e^{(2\xi-\xi_{p})\frac{v}{u}}\sum_{N\geq \a}C_{12}^{p}u^{N-\a}v^{\a}|N,\a\>_p,
\ee
where the state
\bea
|N,\a\>_p&=&\sum_{\stackrel{ \{\vec{k},\vec{q}\},}{K+Q=N,\,\a\leq N}
}\b_{12}^{p\{\vec{k},\vec{q}\},\a}L_{\vec{k}}M_{\vec{q}}|\D_p,\xi_p\>,
\eea
is a descendant state at level $N$ in the BMS module,
\be
L_{0}|N,\a\>_p=(\D_{p}+N)|N,\a\>_p.
\ee
We now act with the generators $L_{n>0}$ on both sides of sides of equation \eqref{ope} and demand that they should transform in the same way.
On the LHS, we have 
\bea
&&L_{n}\phi_{1}(u,v)|\D,\xi\> = [L_{n},\phi_{1}(u,v)]|\D,\xi\>\cr
& &= [u^{n+1}\p_{u}+(n+1)u^{n}v\p_{v}+(n+1)(\D u^{n}-n\xi u^{n-1}v)]\phi_{1}(u,v)|\D,\xi\>.
\eea
Substituting the RHS of \eqref{ope} in the above equation, we have
\bea
 &  & \sum_{p}C_{p}^{12}u^{-2\D+\D_{p}}\,e^{(2\xi-\xi_{p})\frac{v}{u}}L_{n}\sum_{N,\a}u^{N-\a}v^{\a}|N,i\>_p\cr
 & = & \sum_{p}C_{p}^{12}u^{-2\D+\D_{p}}\,e^{(2\xi-\xi_{p})\frac{v}{u}}\cr
 &  & \sum_{N,\a}u^{N-\a +n}v^{\a}\left(N+n \a -\D +n \D +\D _p\right)|N,\a\>_p\cr
 &  & +u^{N-\a -1}v^{\a +1}\left(n \xi -n^2 \xi -n \xi _p\right)|N,\a\>_p.
\eea
If we equate the coefficients of $u^{-2\D+\D_{p}}\,e^{(2\xi-\xi_{p})\frac{v}{u}}u^{K+n-\a}v^{\a}$
on both sides, we get the recursion relation
\newpage
\bea
L_{n}|N+n,\a\>_p & = & \left(N+n \a -\D +n \D +\D _p\right)|N,\a\>_p \cr
 &  & +\left(n \xi -n^2 \xi -n \xi _p\right)|N,\a -1\>_p.
 \label{recurL}
\eea
Similarly, demanding that both sides of the OPE transform the same way under $M_0$ and $M_{n>0}$ we get two more recursion relation 
\bea
M_{0}|N,\a\>_p = \xi_{p} |N,\a,\>_p-(\a +1)|N,\a +1\>_p,
 \label{recurM0}
\eea
\bea
M_{n}|N+n,\a\>_p  =  \left((n-1) \xi +\xi _p\right)|N,\a\>_p-(\a+1)|N,\a+1\>_p.
\label{recurM}
\eea
These three recursion relations can be used to find all the coefficients $\b_{12}^{p\{\vec{k},\vec{q}\},\a}$. We have shown this calculation for level 1 and level 2 in the next section.

\subsection{Finding the coefficients}

At level zero we have
\be
|N=0,\a=0\>_p=\b_{12}^{p\{0,0\},0}|\D_{p},\xi_{p}\>=|\D_{p},\xi_{p}\>.
\ee

\subsubsection*{Level 1}
The states in level 1 are given by 
\be
|1,\a\>_p=\b_{12}^{p\{1,0\},\a}L_{-1}|\D_{p},\xi_{p}\>+\b_{12}^{p\{0,1\},\a}M_{-1}|\D_{p},\xi_{p}\>,\,\,\a=0,1.
\ee
First let us note that
\bea
M_0|1,\a\>_p &=& \xi_p \b_{12}^{p\{1,0\},\a} L_{-1}|\D_{p},\xi_{p}\> + \left(\b_{12}^{p\{1,0\},\a} + \xi_p \b_{12}^{p\{0,1\},\a}\right)M_{-1}|\D_{p},\xi_{p}\>,\\
M_1|1,\a\>_p &=& 2\xi_p \b_{12}^{p\{1,0\},\a}|\D_{p},\xi_{p}\>, \\
L_1|1,\a\>_p &=& 2\left(\D_p \b_{12}^{p\{1,0\},\a}+ \xi_p \b_{12}^{p\{0,1\},\a}\right)|\D_{p},\xi_{p}\>.
\eea
\bigskip
\begin{table}[ht]
\def\arraystretch{1.5}
\centering
\begin{tabular}{ |c|c| } 
 \hline
$\b_{12}^{p\{1,0\},0}=\frac{1}{2}$  & $\b_{12}^{p\{0,1\},0}=0$ \\
\hline
$\b_{12}^{p\{1,0\},1}=0$  & $\b_{12}^{p\{0,1\},1}=-\frac{1}{2}$ \\
\hline
\end{tabular}
\caption{Coefficients of OPE at level 1.}
\label{level1}
\end{table}
\bigskip

\noindent
Then using the recursion relation \eqref{recurM0}, we have
\be
M_{0}|1,1\>_p  =  \xi_{p}|1,1\>_p \implies \b_{12}^{p\{1,0\},1}M_{-1}|\D_{p},\xi_{p}\>  =  0,
\ee
\be
M_{0}|1,0\>_p  =  \xi_{p}|1,0\>_p-|1,1\>_p \implies \left(\b_{12}^{p\{1,0\},0}+\b_{12}^{p\{0,1\},1}\right) M_{-1}|\D_{p},\xi_{p}\> = 0,
\ee
giving us
\be
\b_{12}^{p\{1,0\},1}=0,\,\,\,\,\b_{12}^{p\{1,0\},0}=-\b_{12}^{p\{0,1\},1}.
\ee
Now, using the recursion relation \eqref{recurM} with $N=0,n=1,\a=0$, we have
\be
M_{1}|1,0\>_p  =  \xi_{p}|\D_{p},\xi_{p}\>\implies  \xi_p\left(2\b_{12}^{p\{1,0\},0} -1\right)|\D_{p},\xi_{p}\>=0,
\ee
giving us the coefficients
\be
\b_{12}^{p\{1,0\},0}=\frac{1}{2},\,\,\,\b_{12}^{p\{0,1\},1}=-\frac{1}{2}.
\ee
With $N=0,n=1,\a=0$, \eqref{recurL} gives the recursion relation 
\be
L_{1}|1,0\>_p  =  \D_{p}|\D_{p},\xi_{p}\>\implies 2\b_{12}^{p\{0,1\},0}\xi_{p}|\D_{p},\xi_{p}\> = 0,
\ee
giving us 
\be
\b_{12}^{p\{0,1\},0}=0.
\ee
The various coefficients are collected above in Table \refb{level1}. We can see that these match with the coefficients in (A.7) of \cite{Bagchi:2009pe}.

\subsubsection*{Level 2}

The details of the relevant calculations at level 2 are presented in Appendix A. We collect all these coefficients in Table \eqref{level2}.
\begin{table}[ht]
\def\arraystretch{1.7}
\centering
\begin{tabular}{|c|c|c|} 
\hline
$\b_{12}^{p\{2,0\},0}=\frac{1}{8}$ & $\b_{12}^{p\{(0,1),0\},0}=\frac{4\xi+\xi_{p}}{8(3c_M+2\xi_{p})}$ & $\b_{12}^{p\{1,1\},0}=-\frac{12 \xi -6 c_M-\xi _p}{16 \xi _p(3 c_M+2 \xi _p)}$\\
\hline
\multicolumn{3}{|c|}{
$\b_{12}^{p\{0,2\},0} = \frac{-36 c_M^2 \left(1+\D _p\right)+24 c_M \left(3 \xi +\D _p \left(3 \xi -2 \xi _p\right)+(1-3 \D ) \xi _p\right)+\xi _p \left(-60 \xi +\D _p \left(96 \xi -4 \xi _p\right)+5 \xi _p-48 \D  \xi _p+18 c_L \left(4 \xi +\xi _p\right)\right)}{64 \xi _p^2 \left(3 c_M+2 \xi _p\right)^2}$} \\
\hline
\multicolumn{3}{|c|}
{$\b_{12}^{p\{0,(0,1)\},0}=\frac{36 \xi -24 \xi  c_L-18 c_M+24 \D  c_M-16 \xi  \D _p+6 c_M \D _p-3 \xi _p+16 \D  \xi _p-6 c_L \xi _p}{16 \left(3 c_M+2 \xi _p\right)^2}$} \\
\hline 
$\b_{12}^{p\{2,0\},1}=0$ & $\b_{12}^{p\{(0,1),0\},1}=0$ & $\b_{12}^{p\{1,1\},1}=-\frac{1}{4}$ \\
\hline
\multicolumn{2}{|c|}
{$\b_{12}^{p\{0,2\},1}=\frac{12 \xi -6 c_M-\xi _p}{16 \xi _p(3 c_M+2 \xi _p)}$} &  $\b_{12}^{p\{0,(0,1)\},1}=-\frac{4\xi+\xi_{p}}{4(3c_M+2\xi_{p})}$ \\
\hline
$\b_{12}^{p\{2,0\},2}=0$ & $\b_{12}^{p\{(0,1),0\},2}=0$ & $\b_{12}^{p\{1,1\},2}=0$ \\
\hline
\multicolumn{2}{|c|}
{$\b_{12}^{p\{0,2\},2}=0$}  & $\b_{12}^{p\{0,(0,1)\},2}=\frac{1}{8}$\\
\hline
\end{tabular}
\caption{Coefficients of OPE at level 2.}
\label{level2}
\end{table}

\newpage

\subsection{BMS blocks, crossing symmetry and bootstrap}
We have seen that BMS-invariant theories are completely specified by the structure constants, the spectrum of primary fields, and the central charge. However any given sets of these inputs need not always constitute a consistent theory; they have to satisfy an infinite set of equations analogous to the conformal case which we will call the BMS bootstrap equation. This equation comes from self consistency of the OPE, namely that it has to be associative when applying inside the correlator. More precisely, if we use the OPE inside the correlator, the resulting correlator should not depend on which two neighbouring primary operators we applied the OPE. We will study this requirement by considering the four-point function, which for a BMS-invariant theory has the structure  
\bea
\<\prod_{i=1}^{4}\phi_{i}(u_{i},v_{i})\>&=&\prod_{1\leq i<j\leq4}u_{ij}^{\sum_{k=1}^{4}\D_{ijk}/3}e^{-\frac{v_{ij}}{u_{ij}}\sum_{k=1}^{4}\xi_{ijk}/3} F_{BMS}(u,v)\cr
&\equiv& P(\{\D_i,\xi_i,u_{ij},v_{ij}\})F_{BMS}(u,v)
\label{eqn:4pt}
\eea
where the BMS analogues of the cross ratio $u$ and $v$ given by
\be
u=\frac{u_{12}u_{34}}{u_{13}u_{24}},\,\,\,\frac{v}{u}=\frac{v_{12}}{u_{12}}+\frac{v_{34}}{u_{34}}-\frac{v_{13}}{u_{13}}-\frac{v_{24}}{u_{24}}
\ee 
are invariant under the global coordinate transformation generated by $L_{0,\pm1},M_{0,\pm1}$. We can conveniently do a global coordinate transformation such that 
\be   
\{(u_i,v_i) \} \rightarrow \{(\infty,0), (1,0), (u,v), (0,0)\},
\label{eqn:gt}
\ee 
where $i=1,...,4$. Correspondingly, we define
\be
\lim_{u_{1}\rightarrow\infty,v_{1}\rightarrow0}u_1^{2\D_{1}}\exp\left(-\frac{2\xi_{1}v_{1}}{u_{1}}\right)\<\phi_{1}(u_{1},v_{1})\phi_{2}(1,0)\phi_{3}(u,v)\phi_{4}(0,0)\>\equiv G_{34}^{21}(u,v),
\ee
which in terms of the in and out states is given by
\be
G_{34}^{21}(u,v)=\<\D_{1},\xi_{1}|\phi_{2}(1,0)\phi_{3}(u,v)|\D_{4},\xi_{4}\>.
\ee
It can be easily seen that
\be
f(u,v)F_{BMS}(u,v)=G_{34}^{21}(u,v),
\ee
where
\bea
f(u,v)&=&(1-u)^{\frac{1}{3}(\D_{231}+\D_{234})}u^{\frac{1}{3}(\D_{341}+\D_{342})} e^{\frac{v}{3(1-u)}(\xi_{231}+\xi_{234})}e^{-\frac{v}{3u}(\xi_{341}+\xi_{342})}
\eea
So, the four-point function can be expressed in terms of $G_{34}^{21}(u,v)$ as
\be
\<\prod_{i=1}^{4}\phi_{i}(u_{i},v_{i})\> = P(\{\D_i,\xi_i,u_{ij},v_{ij}\}) f(u,v)^{-1}G_{34}^{21}(u,v).
 \label{four_point}
\ee
Now, we may also define
\be
G_{32}^{41}(u,v)=\<\D_{1},\xi_{1}|\phi_{4}(1,0)\phi_{3}(u,v)|\D_{2},\xi_{2}\>,
\ee
and it can be easily seen that these functions  $G_{ij}^{kl}(u,v)$  are related by crossing symmetry
\be
G_{34}^{21}(u,v)=G_{32}^{41}(1-u,-v).
\label{crosssym}
\ee
It is important to emphasise here that the crossing equation that we have obtained above is {\em not the same} as the usual conformal crossing equation \refb{confcross}.  

If we use the OPE between the fields $\phi_3$ and $\phi_4$ in $G_{34}^{21}(u,v)$
we can see that the function can be expressed in terms of the three-point functions of primary fields and their descendants. More precisely, using the OPE, it can be decomposed as  
\be
G_{34}^{21}(u,v)=\sum_{p}C_{34}^{p}C_{12}^{p}A_{34}^{21}(p|u,v),
\ee
where the four-point conformal block $A_{34}^{21}(p|u,v)$ is the sum of all contributions coming from the primary field $\phi_p$ and its descendants and is given by{\footnote{It should be noted that we can only apply the OPE between neighbouring primary fields, so it is understood that the point $(u,v)$ lies between the origin and a circle of radius 1. }}
\bea
A_{34}^{21}(p|u,v)	&=&	(C_{12}^{p})^{-1}u^{-\D_{3}-\D_{4}+\D_{p}}\,e^{(\xi_{3}+\xi_{4}-\xi_{p})\frac{v}{u}}\sum_{N\geq\a}u^{N-\a}v^{\a}\<\D_{1},\xi_{1}|\phi_{2}(1,0)|N,\a\>_p \cr
	&=&	u^{-\D_{3}-\D_{4}+\D_{p}}\,e^{(\xi_{3}+\xi_{4}-\xi_{p})\frac{v}{u}} \cr
	&&	\times\sum_{\{\vec{k},\vec{q}\}}\left(\sum_{\a=0}^{K+Q}\b_{34}^{p\{\vec{k},\vec{q}\},\a}u^{K+Q-\a}v^{\a}\right)\frac{\<\D_{1},\xi_{1}|\phi_{2}(1,0)L_{\vec{k}}M_{\vec{q}}|\D_{p},\xi_{p}\>}{\<\D_{1},\xi_{1}|\phi_{2}(1,0)|\D_{p},\xi_{p}\>} \cr
	&& 
\eea
As we have already seen, the coefficients $\b_{34}^{p\{\vec{k},\vec{q}\},\a}$ can be calculated recursively using BMS symmetry. Thus, the closed form expression of the BMS blocks are completely determined by symmetry and the only dynamical inputs needed to find the four point functions are the structure constants and the spectrum of primary operators appearing in the OPE.    

For the function $G_{32}^{41}(u,v)$ we may use the OPE on $\phi_2$ and $\phi_3$ giving us the expansion
\be
G_{32}^{41}(u,v)=\sum_p C^p_{23}C^{p}_{14} A_{32}^{41}(p|u,v),
\ee
where the blocks $A_{32}^{41}(u,v)$ are given by
\bea 
A_{32}^{41}(p|u,v)&=& u^{-\D_{3}-\D_{2}+\D_{p}}\,e^{(\xi_{3}+\xi_{2}-\xi_{p})\frac{v}{u}} \cr
&&	\times\sum_{\{\vec{k},\vec{q}\}}\left(\sum_{\a=0}^{K+Q}\b_{32}^{p\{\vec{k},\vec{q}\},\a}u^{K+Q-\a}v^{\a}\right)\frac{\<\D_{1},\xi_{1}|\phi_{4}(1,0)L_{\vec{k}}M_{\vec{q}}|\D_{p},\xi_{p}\>}{\<\D_{1},\xi_{1}|\phi_{4}(1,0)|\D_{p},\xi_{p}\>}.\cr
&&
\eea
Now, \eqref{crosssym} must be satisfied, even if we expand both sides using OPE in terms of the BMS blocks, giving us the BMS bootstrap equation 
\be
\boxed{\ \sum_{p}C_{34}^{p}C_{12}^{p}A_{34}^{21}(p|u,v)=\sum_{q}C_{32}^{q}C_{41}^{q}A_{32}^{41}(q|1-u,-v).\ }
\label{eqn:bootstrap}
\ee
This is one of the main initial results of our analysis. 

\medskip

Knowing the BMS blocks, the above equation put a constrain on the structure constants and weights of primary operators in a consistent field theory with BMS symmetry. We can try to solve the bootstrap equation to find all such possible consistent field theories. The only problem is that we do not have a closed form expression of the blocks even though they are fixed by symmetry alone. However, we can find the leading term in a $\frac{1}{c_{L,M}}$ expansion of the blocks. Using this expansion on both sides of \eqref{eqn:bootstrap}, the equation has to be satisfied order by order. The leading order give us the constraint
\be
\sum_{p}C_{34}^{p}C_{12}^{p}g_{34}^{21}(p|u,v)=\sum_{q}C_{32}^{q}C_{41}^{q}g_{32}^{41}(q|1-u,-v),
\ee
where $g_{ij}^{kl}(p|u,v)$ are the large central charge limit of the blocks $A_{ij}^{kl}(p|u,v)$
\be
g_{ij}^{kl}(p|u,v)=\lim_{c_{L,M}\to \infty} A_{ij}^{kl}(p|u,v).
\ee
We will find $g_{ij}^{kl}(p|u,v)$ in the next section. 

\subsection{Differential equations for global blocks from quadratic Casimirs}

For even dimensional CFTs with $d\geq 4$, the closed form expression of the four point conformal blocks was obtained for scalar operators by Dolan and Osborn in \cite{Dolan:2000ut,Dolan:2003hv}. For $2d$ CFTs, their method gives the global conformal blocks, which is the large central charge limit of the full Virasoro conformal blocks, as we have mentioned in the previous section. In this section we will employ this method to obtain the global blocks for BMS algebra, assuming that such a limit will act in a similar manner.

If we take the asymptotic limit $c_L,c_M\rightarrow\infty$ in the OPE \eqref{ope}, \eqref{ope1},  the leading terms are given by the descendant fields generated by $L_{-1}$ and $M_{-1}$. This can be explicitly seen by looking at the coefficients $\beta$ in the limit $c_L, c_M \to \infty$. For levels 1 and 2, this can be verified by the results obtained in previous sections and outlined in Table \refb{level1} and Table \refb{level2}. More precisely, we have
\bea
&& \phi_{3}(u,v)\phi_{4}(0,0)|0,0\> = \cr
&&  \quad \sum_{p,\{k,q\}}u^{-\D_{1}-\D_{2}+\D_{p}}\,e^{(\xi_{3}+\xi_{4}-\xi_{p})\frac{v}{u}}C_{34}^{p}\,\left(\sum_{\a=0}^{N=k+q}\b_{34}^{p\{k,q\},\a}u^{k+q-\a}v^{\a}\right)(L_{-1})^{k}(M_{-1})^{q}|\D_p,\xi_p\>,\cr
&&\quad \quad \quad +\mathcal{O}\left(\frac{1}{c_L},\frac{1}{c_M}\right)+ \ldots ,
\eea
So the function $G_{34}^{21}(u,v)$ has an expansion of the form
\be
\<\D_{1},\xi_{1}|\phi_{2}(1,0)\phi_{3}(u,v)|\D_{4},\xi_{4}\>	= \sum_{p}C_{12}^{p}C_{34}^{p}\,g_{34}^{21}(p|u,v) + \mathcal{O}\left(\frac{1}{c_L},\frac{1}{c_M}\right)+..., 
\ee
where the global block $g_{34}^{21}(p|u,v)$, which is the large central charge limit of  $G_{34}^{21}(u,v)$, is given by
\bea
g_{34}^{21}(p|u,v) &=& u^{-\D_{3}-\D_{4}+\D_{p}}\,e^{(\xi_{3}+\xi_{4}-\xi_{p})\frac{v}{u}} \cr
&&	\times\sum_{\{k,q\}}\left(\sum_{\a=0}^{N=k+q}\b_{34}^{p\{k,q\},\a}u^{N-\a}v^{\a}\right)\frac{\<\D_{1},\xi_{1}|\phi_{2}(1,0)(L_{-1})^{k}(M_{-1})^{q}|\D_{p},\xi_{p}\>}{\<\D_{1},\xi_{1}|\phi_{2}(1,0)|\D_{p},\xi_{p}\>}.\cr
\label{eqn:cb3}
&& 
\eea
More generally, we have
\bea
&&\phi_{3}(u_3,v_3)\phi_{4}(u_4,v_4)|0,0\> = \cr
&&  \quad  \sum_{p,\{k,q\}}u_{34}^{-\D_{1}-\D_{2}+\D_{p}}\,e^{(\xi_{3}+\xi_{4}-\xi_{p})\frac{v_{34}}{u_{34}}}C_{34}^{p}\,\left(\sum_{\a=0}^{N=k+q}\b_{34}^{p\{k,q\},\a}u^{k+q-\a}v^{\a}\right)(L_{-1})^{k}(M_{-1})^{q}\phi_4(u_4,v_4)|0,0\>,\cr
&&\quad \quad \quad  + \mathcal{O}\left(\frac{1}{c_L},\frac{1}{c_M}\right)+...,
\label{eqn:ope3}
\eea
with the four-point function given by the $1/c_{L,M}$ expansion
\be
\<\prod_{i=1}^4\phi_i(u_i,v_i)\>	= \sum_{p}C_{12}^{p}C_{34}^{p}\,\tl{g}_{34}^{21}(p|u,v) + \mathcal{O}\left(\frac{1}{c_L},\frac{1}{c_M}\right)+..., 
\label{4_pt_expand}
\ee
where 
\be
\tl{g}_{34}^{21}(p|u,v)=P(\{\D_i,\xi_i,u_{ij},v_{ij}\})f(u,v)^{-1}g_{34}^{21}(p|u,v).
\label{eqn:rel_blocks}
\ee

It is possible to find the blocks $g_{34}^{21}(p|u,v)$ by demanding that both sides of the OPE transform the same way under the action of the quadratic Casimirs belonging to the global algebra generated by $\{L_{-1},L_0,L_1,M_{-1},M_0,M_1\}$. These Casimirs are given by 
\bea
\mathcal{C}_1 &=& M_0^2-M_{-1}M_1 \\
\mathcal{C}_2 &=& 2L_0M_0-\frac{1}{2}(L_{-1}M_1+L_1M_{-1}+M_1L_{-1}+M_{-1}L_1).
\eea
It can be seen that the states $(L_{-1})^{k}(M_{-1})^{q}\phi_4(u_4,v_4)|0,0\>$ are eigenstates of  $\mathcal{C}_1$ and $\mathcal{C}_2$ since the Casimirs commute with  $L_{-1}$, $M_{-1}$,
\bea
\mathcal{C}_{1,2}(L_{-1})^{k}(M_{-1})^{q}\phi_4(u_4,v_4)|0,0\> &\equiv & \lambda_{1,2}^p (L_{-1})^{k}(M_{-1})^{q}\phi_4(u_4,v_4)|0,0\>\cr,
&& 
\eea
where the eigenvalues are given by
\be
\lambda_1^p =\xi_{p}^{2}, \,\,\,\, \lambda_2^p =	(2\D_{p}\xi_{p}-2\xi_{p}).  
\ee
Consequently, we have
\newpage
\bea
&&\mathcal{C}_{1,2} \phi_{3}(u_3,v_3)\phi_{4}(u_4,v_4)|0,0\> = \cr
&&\sum_{p,\{k,q\}}\lambda_{1,2}^p\,u_{34}^{-\D_{1}-\D_{2}+\D_{p}}\,e^{(\xi_{3}+\xi_{4}-\xi_{p})\frac{v_{34}}{u_{34}}}C_{34}^{p}\,\left(\sum_{\a=0}^{N=k+q}\b_{34}^{p\{k,q\},\a}u^{k+q-\a}v^{\a}\right)(L_{-1})^{k}(M_{-1})^{q}\phi_4(u_4,v_4)|0,0\>\cr
&&\quad \quad \quad + \mathcal{O}\left(\frac{1}{c_L},\frac{1}{c_M}\right)+....
\eea 
After taking the inner product on both sides with $\<\phi_1(u_1,v_1)\phi_2(u_2,v_2)|$, we have
\be 
\<\phi_1(u_1,v_1)\phi_2(u_2,v_2)\mathcal{C}_{1,2}\phi_3(u_3,v_3)\phi_4(u_4,v_4)\> =  \sum_{p} \lambda_{1,2}^p\, C_{12}^{p}C_{34}^{p}\,\tl{g}_{34}^{21}(p|u,v) + \mathcal{O}\left(\frac{1}{c_L},\frac{1}{c_M}\right)+.... 
\ee
On the LHS of the above equation, $\mathcal{C}_{1,2}$ act as differential operators $\mathcal{D}_{1,2}$. More precisely, these differential operators are given by
\bea
&& \mathcal{C}_1\phi_3(y_{3})\phi_4(y_{4})|0\>	\cr
&&=	(M_{0}^{2}-M_{-1}M_{1})\phi_3(y_{3})\phi_4(y_{4})|0\> \cr
&& = [(-u_{3}\p_{v_{3}}+\xi_3-u_{4}\p_{v_{4}} + \xi_4)(-u_{3}\p_{v_{3}}+ \xi_3 - u_{4}\p_{v_{4}} +\xi_4)\cr
&&	-(-\p_{v_{3}}-\p_{v_{4}})(-u_{3}^{2}\p_{v_{3}}+2\xi_3 u_{3}-u_{4}^{2}\p_{v_{4}}+2\xi_4 u_{4})]\phi_3(y_{3})\phi_4(y_{4})|0\> \cr
&& = [2\xi_3 (u_{3}-u_4)\p_{v_{4}}-2\xi_4 (u_3-u_{4})\p_{v_{3}}+(\xi_3+\xi_4)^{2}-(u_3-u_4)^2\p_{v_{3}}\p_{v_{4}}](\phi_3(y_{3})\phi_4(y_{4}))|0\> \cr
&&\equiv  \mathcal{D}_1(\phi_3(y_{3})\phi_4(y_{4}))|0\>,
\eea
\bea
&&\mathcal{C}_2\phi_3(y_{3})\phi_4(y_{4})|0\> \cr
&&=	[2L_0M_0-\frac{1}{2}(L_{-1}M_1+L_1M_{-1}+M_1L_{-1}+M_{-1}L_1)]\phi_3(y_{3})\phi_4(y_{4})|0\> \cr
&&= [2(\D _3+\D _4-1) (\xi _3+\xi _4)+(-2 u_3 \xi _3+2 u_4 \xi _3) \p_{u_4}+(2 u_3 \xi _4-2 u_4 \xi _4) \p_{u_3}\cr
&&+(-2 u_3 \D _4+2 u_4 \D _4+2 v_3 \xi _4-2 v_4 \xi _4) \p_{v_3}
+ (2 u_3 \D _3-2 u_4 \D _3-2 v_3 \xi _3+2 v_4 \xi _3) \p_{v_4}\cr
&&+(u_3^2-2 u_3 u_4+u_4^2) \p_{v_4}\p_{u_3} + (u_3^2-2 u_3 u_4+u_4^2) \p_{v_3}\p_{u_4}\cr
&& + (2 u_3 v_3-2 u_4 v_3-2 u_3 v_4+2 u_4 v_4) \p_{v_3}\p_{v_4}]\phi_3(y_{3})\phi_4(y_{4})|0\>\cr
&&\equiv \mathcal{D}_2(\phi_3(y_{3})\phi_4(y_{4}))|0\>.
\eea
Pulling the differential operator outside the four-point function, we have
\be 
\mathcal{D}_{1,2}\<\phi_1(u_1,v_1)\phi_2(u_2,v_2)\phi_3(u_3,v_3)\phi_4(u_4,v_4)\> =  \sum_{p} \lambda_{1,2}^p\, C_{12}^{p}C_{34}^{p}\,\tl{g}_{34}^{21}(p|u,v) + \mathcal{O}\left(\frac{1}{c_L},\frac{1}{c_M}\right)+.... 
\ee
We then expand the LHS using  \eqref{4_pt_expand}. This gives
\bea
&&\mathcal{D}_{1,2}\sum_{p}C_{12}^{p}C_{34}^{p}\,\tl{g}_{34}^{21}(p|u,v) + \mathcal{O}\left(\frac{1}{c_L},\frac{1}{c_M}\right)+... 
= \sum_{p} \lambda_{1,2}^p\, C_{12}^{p}C_{34}^{p}\,\tl{g}_{34}^{21}(p|u,v) + \mathcal{O}\left(\frac{1}{c_L},\frac{1}{c_M}\right)+.... \cr
&&
\eea
This equation has to be satisfied order by order. The leading order give us a differential equation for 
$\tl{g}_{34}^{21}(p|u,v)$
\bea
\mathcal{D}_{1,2}\sum_{p}C_{12}^{p}C_{34}^{p}\,\tl{g}_{34}^{21}(p|u,v)=\sum_{p} \lambda_{1,2}^p\, C_{12}^{p}C_{34}^{p}\,\tl{g}_{34}^{21}(p|u,v)
\eea
We can decouple this to get differential equations for each block $\tl{g}_{34}^{21}(p|u,v)$,
\be
\mathcal{D}_{\mathcal{C}_{1,2}}\tl{g}_{34}^{21}(p|u,v) = \lambda_{1,2}^p\,\tl{g}_{34}^{21}(p|u,v). 
\ee
Using \eqref{eqn:rel_blocks} we have
\be 
\mathcal{D}_{1,2}\left(P(\{\D_i,\xi_i,u_{ij},v_{ij}\})f(u,v)^{-1}g_{34}^{21}(p|u,v)\right) = \lambda_{1,2}^p\,P(\{\D_i,\xi_i,u_{ij},v_{ij}\})f(u,v)^{-1}g_{34}^{21}(p|u,v).
\ee 
Let us first look at the differential equation associated with $\mathcal{C}_1$, which is given by
\bea
&&\mathcal{D}_1\left(\prod_{1\leq i<j\leq4}e^{-\frac{v_{ij}}{u_{ij}}(\frac{1}{3}\sum_{k=1}^{4}\xi_{k}-\xi_{i}-\xi_{j})}\,f(u,v)^{-1}g_{34}^{21}(p|u,v)\right)\cr
&& = \xi_{p}^{2}\prod_{1\leq i<j\leq4} e^{-\frac{v_{ij}}{u_{ij}}(\frac{1}{3}\sum_{k=1}^{4}\xi_{k}-\xi_{i}-\xi_{j})}\,f(u,v)^{-1}g_{34}^{21}(p|u,v).
\eea
For simplicity let us consider the case where $\D_{i=1,2,3,4}=\D,\,\xi_{i=1,2,3,4}=\xi$. Then the above equation reduces to
\bea
&&\mathcal{D}_1\left(\prod_{1\leq i<j\leq4}e^{\frac{v_{ij}}{u_{ij}}\frac{2\xi}{3}}(1-u)^{\frac{2\D}{3}}u^{\frac{2\D}{3}}e^{\frac{2\xi v}{3(1-u)}}e^{-\frac{2\xi v}{3u}}g_{\D,\xi}(p|u,v)\right)\cr
&&=\xi_{p}^{2}\prod_{1\leq i<j\leq4} e^{\frac{v_{ij}}{u_{ij}}\frac{2\xi}{3}}\,(1-u)^{\frac{2\D}{3}}u^{\frac{2\D}{3}}e^{\frac{2\xi v}{3(1-u)}}e^{-\frac{2\xi v}{3u}}g_{\D,\xi}(p|u,v)),
\eea
where we have used the notation $g_{\D,\xi}(p|u,v)$ for the blocks $g_{34}^{21}(p|u,v)$ in this special case and $\mathcal{D}_1$ is also taken with $\xi_{i=1,2,3,4}=\xi$. If we combine the functions of $u$ and $v$ into 
\be
\hat{g}_{\D,\xi}(p|u,v)=(1-u)^{\frac{2\D}{3}}u^{\frac{2\D}{3}}e^{\frac{2\xi v}{3(1-u)}}e^{-\frac{2\xi v}{3u}}g_{\D,\xi}(p|u,v),
\ee
then we have
\bea
\mathcal{D}_1\left(\prod_{1\leq i<j\leq4}e^{\frac{v_{ij}}{u_{ij}}\frac{2\xi}{3}}\G_{\D,\xi}(p|u,v)\right) = \xi_{p}^{2}\prod_{1\leq i<j\leq4} e^{\frac{v_{ij}}{u_{ij}}\frac{2\xi}{3}}\,\hat{g}_{\D,\xi}(p|u,v),
\eea
which explicitly is given by
\bea
&&\Big[4 \xi ^2-\xi _p^2 +\frac{4}{3} \xi ^2 (u_{31}^{-1}+u_{32}^{-1}+u_{34}^{-1}) u_{43} -\frac{4}{3} \xi ^2 (u_{41}^{-1}+u_{42}^{-1}+u_{43}^{-1})u_{43}\cr
&& -\frac{4}{9} \xi ^2 (u_{31}^{-1}+u_{32}^{-1}+u_{34}^{-1})(u_{41}^{-1}+u_{42}^{-1}+u_{43}^{-1})u_{34}^2
+\left(2 \xi u_{43}+\frac{2}{3} \xi u_{34}u_{43}(u_{41}^{-1}+u_{42}^{-1}+u_{43}^{-1})\right)\p_{v_3}\cr
&& + \left (-2 \xi u_{43}+\frac{2}{3} \xi (u_{31}^{-1}+u_{32}^{-1}+u_{34}^{-1})(u_{34}u_{43}\right)\p_{v_4} 
-u_{34}^2 \p_{v_3}\p_{v_4}\Big]\hat{g}_{\D,\xi}(p|u,v)=0,
\eea
where $u_{ij}=u_i-u_j$.
Under the global conformal transformation \eqref{eqn:gt}, the above equation reduces to
\bea
\left[4 (u-2)^2 \xi ^2+9 (u-1) \xi _p^2 -12 u (2-3 u+u^2) \xi  \p_v 
 +9 (u-1)^2 u^2\p_v^2\right]\hat{g}_{\D,\xi}(p|u,v)=0. 
\eea
In terms of the global blocks $g_{\D,\xi}(p|u,v)$, the above equation is given by
\be
[(4 (u-1) \xi ^2+\xi _p^2) -4\xi u(u-1)  \p_v + u^2(u-1) \p_v^2 ]g_{\D,\xi}(p|u,v)=0.
\ee
The differential equation gets simpler if we introduce a function
\be
k(p|u,v)=u^{2\D}e^{-\frac{2\xi v}{u}}g_{\D,\xi}(p|u,v).
\ee 
Plugging this back into the above equation, we have the simplified version: 
\be
\left[\p_v^2+\frac{\xi_p^2}{u^2(u-1)}\right]h(p|u,v)=0.
\label{diffeqn1}
\ee
Now let us look at the differential equation associated with $\mathcal{C}_2$. For simplicity we again only consider the case where $\D_{i=1,2,3,4}=\D,\,\xi_{i=1,2,3,4}=\xi$. We have,
\be
\mathcal{D}_2\left(\prod_{1\leq i<j\leq4}u_{ij}^{-\frac{2\D}{3}}e^{\frac{v_{ij}}{u_{ij}}\frac{2\xi}{3}}\,\hat{g}_{\D_{p},\xi_{p}}(u,v))\right) = (2\D_p\xi_p-2\xi_p)\prod_{1\leq i<j\leq4}u_{ij}^{-\frac{2\D}{3}}e^{\frac{v_{ij}}{u_{ij}}\frac{2\xi}{3}}\,\hat{g}_{\D,\xi}(p|u,v).
\ee
Under the global conformal transformation \eqref{eqn:gt}, the above differential equation reduces to
\bea
&&[(2 \xi (-6+8 \D -2 u^3 \D -2 u (-6+8 \D +v \xi )+u^2 (-6+10 \D +v \xi ))\cr
&&-9 (-1+u)^2(-1+\D _p) \xi _p) - 12 (-2+u) (-1+u)^2 u \xi \p_u \cr 
&&+ 3 (-1+u)^2 (u^2 (6+4 \D ) +8 v \xi -8 u (\D +v \xi ))\p_v  + 18 (-1+u)^3 u^2 \p_v\p_u \cr
&& + 9 (-1+u)^2 u (-2+3 u) v\p_v^2 ]\,\hat{g}_{\D,\xi}(p|u,v)=0. 
\eea
In terms of the global blocks $g_{\D,\xi}(p|u,v)=(1-u)^{-\frac{2\D}{3}}u^{-\frac{2\D}{3}}e^{-\frac{2\xi v}{3(1-u)}}e^{\frac{2\xi v}{3u}}\hat{g}_{\D,\xi}(p|u,v)$, we have
\bea
&&[2(2 \xi  (-1+2 \D -2 u \D +v \xi )-(-1+\D _p) \xi _p)+2(u^2 (1+2 \D )+2 v \xi -2 u (\D +2 v \xi ))\p_v \cr
&&+u (-2+3 u) v \p_v^2-4 (-1+u) u \xi \p_u + 2 (-1+u) u^2 \p_v\p_u]g_{\D,\xi}(p|u,v)=0.
\eea
In terms of the function $k(p|u,v)$, the differential equation again gets simpler
\bea
\left[ u^2 \p_v - (1-\frac{3}{2} u) u v \p_v^2 + (u-1) u^2 \p_u\p_v \right] k(p|u,v) =(\D _p - 1) \xi _p \ k(p|u,v). 
\label{diffeqn2}
\eea
\bigskip

\subsection{Solution of the BMS Global block}
In this subsection, we find the explicit solution for the differential equation for the global BMS block. 
The general solutions of the above differential equation \refb{diffeqn1} are given by
\be
k_1(p|u,v)=A_{\D_{p},\xi_{p}}(u)e^{\frac{\xi_p}{u\sqrt{1-u}}v},\,\,k_2(p|u,v)=B_{\D_{p},\xi_{p}}(u)e^{-\frac{\xi_p}{t\sqrt{1-u}}v}.
\label{sol:diff1}
\ee
Substituting \eqref{sol:diff1} in the second differential equation \eqref{diffeqn2} we have
\bea
&&A_{\D_p,\xi_p}(u)(-2(1+\sqrt{1-u})+(2+\sqrt{1-u}) u-2 (-1+u) \Delta _p)\cr
&&+2 (1-u)^{3/2} u\frac{dA_{\D_p,\xi_p}(u)}{du}=0,
\eea
\bea
&&B_{\D_p,\xi_p}(u)(2-2 \sqrt{1-u}+(-2+\sqrt{1-u}) u+2 (-1+u) \Delta _p)\cr
&&+2 (1-u)^{3/2} u\frac{dB_{\D_p,\xi_p}(u)}{du}=0. 
\eea
The solutions of the above differential equations are given by
\be
A_{\D_p,\xi_p}(u)=K_A\,\frac{u^{\D _p} (1-\sqrt{1-u})^{2-2\D _p}}{\sqrt{1-u}},\,\,\,
B_{\D_p,\xi_p}(u)=K_B\,\frac{u^{\D _p} (1+\sqrt{1-u})^{2-2\D _p}}{\sqrt{1-u}},
\ee
where $K_A$ and $K_B$ are constant of integration.
So the most general solution of $h(p|u,v)$ is given by
\be
k(p|u,v)=K_A\,\frac{u^{\D _p} (1-\sqrt{1-u})^{2-2\D _p}}{\sqrt{1-u}}e^{\frac{\xi_p}{u\sqrt{1-u}}v}+K_B\,\frac{u^{\D _p} (1+\sqrt{1-u})^{2-2\D _p}}{\sqrt{1-u}}e^{-\frac{\xi_p}{u\sqrt{1-u}}v}.
\label{full_sol}
\ee
Note again that the blocks are defined only for $u^2+v^2<1$. So we don't have to consider the case $u>1$, where  the above equation becomes oscillatory.

Now, we need boundary conditions to find the constant of integration. Looking at \eqref{eqn:cb3}, we can see that $k(p|u,v)$ is given by
\bea
k(p|u,v) &=& u^{\D_{p}}\,e^{-\xi_p\frac{v}{u}} \cr
&&	\times\sum_{\{k,q\}}\left(\sum_{\a=0}^{N=k+q}\b_{34}^{p\{k,q\},\a}u^{N-\a}v^{\a}\right)\frac{\<\D,\xi|\phi(1,0)(L_{-1})^{k}(M_{-1})^{q}|\D_{p},\xi_{p}\>}{\<\D,\xi|\phi(1,0)|\D_{p},\xi_{p}\>}.\cr
&& 
\eea
Let us show a few of the terms in the summation. We know that $\b_{12}^{p\{0,0\},0}=1$, $\b_{12}^{p\{1,0\},0}=\frac{1}{2}$, $\b_{12}^{p\{0,1\},0}=-\frac{1}{2}$ and 
\be
\frac{\<\D,\xi|\phi(1,0)L_{-1}|\D_{p},\xi_{p}\>}{\<\D,\xi|\phi(1,0)|\D_{p},\xi_{p}\>}=\D_p,\,\,\frac{\<\D,\xi|\phi(1,0)M_{-1}|\D_{p},\xi_{p}\>}{\<\D,\xi|\phi(1,0)|\D_{p},\xi_{p}\>}=\xi_p.
\ee
So, we have
\be
k(p|u,v) = u^{\D_{p}}\,e^{-\xi_p\frac{v}{u}}\left(1+\frac{\D_p}{2}u-\frac{\xi_p}{2}v+...\right).  
\label{cb:expand}
\ee
Expanding our solution \eqref{full_sol} and comparing with the above equation we can find the values $K_A$ and $K_B$. For $|u|<1$, the expansion of our solution is given by
\bea
k(p|u,v)&=& K_Au^{\D _p}e^{-\frac{\xi_p v}{u}}2^{2-2 \D _p}u^{-2 \Delta _p}\left(2^{-2+2 \Delta _p} u^22^{-1+2 \Delta _p} \xi _p uv+2^{-1+2 \Delta _p} \xi _p^2v^2\right) \cr
&&+ K_B\,u^{\D _p}e^{-\frac{\xi_p v}{u}}2^{2-2 \D _p}\left(1-\frac{1}{2}\xi _p v+\frac{1}{2}\D _p u+...\right).
\eea
Comparing this with \eqref{cb:expand}, we have
\be
K_A=0,\,\,K_B=2^{2 \D _p-2}.
\ee
So, we have
\bea{}
g_{\D,\xi}(p|u,v)= 2^{2 \D _p-2}\, \left(1-u\right)^{-1/2}  \exp{\left(\frac{-\xi_p v}{u\sqrt{1-u}} +2\xi \frac{v}{u} \right)} &&u^{\D _p-2\D} (1+\sqrt{1-u})^{2-2\D _p}, \cr
&& \mbox{where} \quad |u|<1.
\label{BMS_global_blocks}
\eea
This is the explicit form of the global BMS blocks in the limit of large central charges and is one of the main results of our initial analysis. 

\newpage

\section{BMS Bootstrap: the limiting analysis}
In this section, we will consider the limit of the 2d conformal algebra that leads to the BMS$_3$ algebra, or equivalently, the 2d GCA. There are two distinct limits that do this; one can be looked upon as a non-relativistic limit and the other as an ultra-relativistic limit. We shall first discuss them and then focus on the non-relativistic limit. We shall then proceed to reproduce some of the answers obtained in the previous section in the light of this limit. 

\subsection{Two contractions of 2d conformal algebra}
There are two distinct contractions of two copies of the relativistic Virasoro algebra that lead to the 2d GCA. At an algebraic level these are given by
\be\label{nr}
L_n = \L_n + \bcL_n, \quad M_n = \e \left(\L_n - \bcL_n \right) \quad \Rightarrow \quad \mbox{Non-Relativistic Limit.}
\ee
and
\be\label{ur}
L_n = \L_n + \bcL_{-n}, \quad M_n = \e \left(\L_n + \bcL_{-n} \right) \quad \Rightarrow \quad \mbox{Ultra-Relativistic Limit.}
\ee
To see why these contractions are so named, it is instructive to look at the generators of the Virasoro algebra on the cylinder and follow the contraction. 
The conformal generators on the cylinder are
\be{}
\L_n = e^{in\w} \p_\w, \quad \bcL_n = e^{in\bw} \p_{\bw}, \quad \w, \bw = \t \pm \s.
\ee
Now in the non-relativistic limit, the co-ordinates on the cylinder scale as $(\s, \t) \to (\e\s,\t)$. It is now clear that if you take \refb{nr}, in this limit this combination gives well-behaved vector fields
\be\label{nrvec}
L_n = e^{in\t}(\p_\t+in\t \p_\s), \quad M_n = e^{in\t} \p_\s.
\ee
These close to form the algebra \refb{gca2d}. The ultra-relativistic limit is when we scale the co-ordinates on the cylinder as $(\s, \t) \to (\s,\e\t)$. It is easy to check that using \refb{ur}, one now gets well defined vector fields
\be\label{urvec}
L_n = e^{in\s}(\p_\s+in\t \p_\t), \quad M_n = e^{in\s} \p_\t.
\ee
It is gratifying to see that \refb{nrvec} and \refb{urvec} are related by a swap of $\s \leftrightarrow \t$, as one would expect. In a combined notation, we can write
\be\label{combvec}
L_n = e^{inU}(\p_U+in\t \p_V), \quad M_n = e^{inU} \p_V
\ee
where U is the un-contracted direction and V is the contracted direction in the field theory. We are interested in the ``plane'' representations. The mapping between the two representations is given by
\be\label{pcmap}
u = e^{iU}, \quad v=iV e^{iU}.
\ee
This connects us to the notation of the previous section. In particular \refb{combvec} goes to \refb{planevec} under the map \refb{pcmap}.  

\paragraph{Highest weight representations and limits:} Throughout the previous section on the intrinsic analysis of the construction of the bootstrap equation and the solution of the BMS block in the limit of large central charges, we have worked in the highest weight representation, as was done in the case of the relativistic CFT analysis mentioned previously. We will now attempt to understand some aspects of this through the limit. We seem to have two distinct limits \refb{nr}, \refb{ur} and hence two distinct ways of achieving this. 

This is, however, not true. Let us consider the ultra-relativistic limit \refb{ur}. We clearly see that there is a mixture of positive and negative modes of the Virasoro algebra in this limit. The Virasoro highest weight condition \refb{hwV} thus does not reduce to the BMS highest weight condition \refb{hwB}. The Virasoro highest weight condition leads to a distinct class of representations called induced representations \cite{Barnich:2014kra, Barnich:2015uva, Campoleoni:2016vsh}. 

The non-relativistic limit \refb{nr} is conducive to reproducing our earlier results, as in this case there is no mixing of positive and negative modes and the Virasoro highest weight representations do go over the the BMS highest weight representations. We will thus be focusing on the non-relativistic limit in an attempt to reconstruct the answers previously obtained from the intrinsic analysis. 

\subsection{Reproducing coefficients of OPE}
We will start by attempting to recover the coefficients of the BMS OPE that we obtained in the previous section through the non-relativistic limit. 

For simplicity take $h_{1}=h_{2}=h,\,\,\h_{1}=\h_{2}=\h$. Acting \eqref{opeCFT} on the vacuum, the RHS is given by
\be
\sum_{p,\{\vec{k},\vec{\bar{k}}\}}\frak{C}_{p12} \ \mathcal{B}_{12}^{p\{\vec{k}\}}\bar{\mathcal{B}}_{12}^{p\{\vec{\bar{k}}\}}z^{h_{p}-2h+K}\,\z^{\h_{p}-2\h+\bar{K}}\L{}_{\vec{k}}\bcL_{\vec{\bar{k}}}|h_{p},\h_{p}\>.
\label{opeCFT1}
\ee
In terms of the space time co-ordinates $z=t+x$, $\z=t-x$. We will consider the non-relativistic contraction
\be
t\rightarrow t,\,\,\,\,x\rightarrow\e x.
\label{contraction}
\ee
The reason for doing so, and not considering the ultra-relativistic contraction $(t\to\e t, x\to x)$, has already been stated above.  
We note again that in our notation in the previous section $v$ is the co-ordinate which is contracted. In this case \eqref{opeCFT1} is given by
\bea
 &  & \sum_{p,\{\vec{k},\vec{\bar{k}}\}}\frak{C}_{p12} \ \mathcal{B}_{12}^{p\{\vec{k}\}}\bar{\mathcal{B}}_{12}^{p\{\vec{\bar{k}}\}}(t+\e x)^{h_{p}-2h+K}\,(t-\e x)^{\h_{p}-2\h+\bar{K}}\,\L_{\vec{k}}\bcL_{\vec{\bar{k}}}|h_{p},\h_{p}\>\cr
 & = & \sum_{p,\{\vec{k},\vec{\bar{k}}\}}\frak{C}_{p12} \ \mathcal{B}_{12}^{p\{\vec{k}\}}\bar{\mathcal{B}}_{12}^{p\{\vec{\bar{k}}\}}t^{h_{p}-2h+K+\h_{p}-2\h+\bar{K}}\left(1+\e\frac{x}{t}\right)^{h_{p}-2h+K}\,\left(1-\e\frac{x}{t}\right)^{\h_{p}-2\h+\bar{K}}\,\L_{\vec{k}}\bcL_{\vec{\bar{k}}}|h_{p},\h_{p}\>\cr
 & = & \sum_{p,\{\vec{k},\vec{\bar{k}}\}}\frak{C}_{p12} \ \mathcal{B}_{12}^{p\{\vec{k}\}}\bar{\mathcal{B}}_{12}^{p\{\vec{\bar{k}}\}}t^{h_{p}+\h_{p}-2(h+\h)+K+\bar{K}}\exp\left(\log\left((1+\e\frac{x}{t})^{h_{p}-2h}\,(1-\e\frac{x}{t})^{\h_{p}-2\h}\right)\right)\cr
 &  & \quad\quad \times \left(1+\e\frac{x}{t}\right)^{K}\,\left(1-\e\frac{x}{t}\right)^{\bar{K}}\,\L_{\vec{k}}\bcL_{\vec{\bar{k}}}|h_{p},\h_{p}\>\cr
 & = & \sum_{p,\{\vec{k},\vec{\bar{k}}\}}\frak{C}_{p12} \ \mathcal{B}_{12}^{p\{\vec{k}\}}\bar{\mathcal{B}}_{12}^{p\{\vec{\bar{k}}\}}t^{h_{p}+\h_{p}-2(h+\h)+K+\bar{K}}\exp\left(\left(h_{p}-\h_{p}-2(h-\h)\right)\left(\e\frac{x}{t}+\mathcal{O}(\e^{2})\right)\right)\cr
 &  & \quad \quad \times \left(1+\e\frac{x}{t}\right)^{K}\,\left(1-\e\frac{x}{t}\right)^{\bar{K}}\,\L_{\vec{k}}\bcL_{\vec{\bar{k}}}|h_{p},\h_{p}\>.
 \label{opecontrac}
\eea
Taking the non-relativistic limit $\e\rightarrow0$, we have
\be
\D=\lim_{\e\rightarrow0}(h+\h),\,\,\,\,\xi=\lim_{\e\rightarrow0}\e(\h-h),\,\,\,\L_{n}=\frac{1}{2}(L_{n}-\frac{{1}}{\e}M_{n}),\,\,\,\bcL_{n}=\frac{1}{2}(L_{n}+\frac{{1}}{\e}M_{n}).
\ee
Putting these in equation \eqref{opecontrac}, we get
\bea
 &  & \sum_{p,\{\vec{k},\vec{\bar{k}}\}}\frak{C}_{p12} \ \mathcal{B}_{12}^{p\{\vec{k}\}}\bar{\mathcal{B}}_{12}^{p\{\vec{\bar{k}}\}}t^{\D-2\D_{p}}t^{K+\bar{K}}\exp\left(2\xi-\xi_{p}+\mathcal{O}(\e^{2})\right)\cr
 &  & \quad \quad \left(1+\e\frac{x}{t}\right)^{K}\,\left(1-\e\frac{x}{t}\right)^{\bar{K}}\,\frac{1}{2}\left(L-\frac{{1}}{\e}M\right)_{\vec{k}}\frac{1}{2}\left(L+\frac{{1}}{\e}M\right)_{\vec{\bar{k}}}|\D_{p},\xi_{p}\>.
 \label{ope_nrlimit}
\eea

\subsection*{Level 1}
If we look at only the level one states in \eqref{ope_nrlimit} without the common factor $\frak{C}_{p12} \ t^{\D-2\D_{p}}\exp\left(2\xi-\xi_{p}\right)$, we have
\bea
 &  & t\left(\mathcal{B}_{12}^{p\{1\}}\left(1+\e\frac{x}{t}\right)\frac{1}{2}\left(L_{-1}-\frac{{1}}{\e}M_{-1}\right)+\mathcal{B}_{12}^{p\{\bar{1}\}}\left(1-\e\frac{x}{t}\right)\frac{1}{2}\left(L_{-1}+\frac{{1}}{\e}M_{-1}\right)\right)|\D_{p},\xi_{p}\>\cr
 & = & \left(\frac{t}{2}\left(\mathcal{B}_{12}^{p\{1\}}+\mathcal{B}_{12}^{p\{\bar{1}\}}\right)L_{-1}-\frac{x}{2}\left(\mathcal{B}_{12}^{p\{1\}}+\mathcal{B}_{12}^{p\{\bar{1}\}}\right)M_{-1}+\mathcal{O}\left(\e^{2}\right)\right)|\D_{p},\xi_{p}\>.
\eea
Using the known coefficients of CFT $\mathcal{B}_{12}^{p\{1\}}=\mathcal{B}_{12}^{p\{\bar{1}\}}=\frac{1}{2}$,
we can see that
\be
\b_{12}^{p\{1,0\},0}=\frac{1}{2},\,\,\,\b_{12}^{p\{0,1\},1}=-\frac{1}{2},\,\,\,\b_{12}^{p\{1,0\},1}=\b_{12}^{p\{0,1\},0}=0.
\ee
These match with the coefficients in Table \eqref{level1}. 
\subsection*{Level 2}
The details of the level 2 calculations are presented in Appendix B. We see that the coefficients again match up with the answers previously obtained in the intrinsic method. 

\medskip

\subsection{Differential equation for blocks from the limiting case}
Having demonstrated that the coefficients of the BMS OPE can be recovered from a limit of the OPE for the Virasoro algebra, we now go on to demonstrate that some other key features of our intrinsic analysis can also be reproduced in this limit. In this subsection, we concentrate on deriving the differential equations for the global BMS blocks from the corresponding equations for the global CFT blocks. 

Under the contraction \eqref{contraction} and the definition in \eqref{hw}, the differential equations in  \eqref{diff_eqn_cft} transforms to
\bea
&&\left[(t+\e x)^2(1-(t+\e x))\frac{1}{4}(\p_t+\frac{1}{\e}\p_x)^2-(t+\e x)^2 \frac{1}{2}(\p_t+\frac{1}{\e}\p_x) \right.\cr
&&\quad\quad\quad \left. - \frac{1}{2}(\Delta_p -\frac{\xi_p }{\e})( \frac{1}{2}(\Delta_p -\frac{\xi_p }{\e})+1)\right]\mathcal{K}_{\D_p,\xi_p}(t,x)=0, 
\eea
\bea
&&\left[(t-\e x)^2(1-(t-\e x))\frac{1}{4}(\p_t-\frac{1}{\e}\p_x)^2-(t-\e x)^2 \frac{1}{2}(\p_t-\frac{1}{\e}\p_x) \right.\cr
&&\quad\quad\quad \left. - \frac{1}{2}(\Delta_p +\frac{\xi_p }{\e})( \frac{1}{2}(\Delta_p +\frac{\xi_p }{\e})+1)\right]\mathcal{K}_{\D_p,\xi_p}(t,x)=0. 
\eea
Taking the limit $\e\rightarrow 0$, the only remaining part in both differential equations is
\be
\left[\p_x^2+\frac{\xi_p^2}{t^2(t-1)}\right]\mathcal{K}_{\D_p,\xi_p}(t,x)=0.
\ee
We can also subtract one differential equation from the other and then take the limit which gives rise to the differential equation
\bea
\left[ t^2 \p_x - (1-\frac{3}{2} t) t x \p_x^2 + (t-1) t^2 \p_t\p_x \right] \mathcal{K}_{\D_p,\xi_p}(t,x) =(\D _p - 1) \xi _p \ \mathcal{K}_{\D_p,\xi_p}(t,x). 
\eea
These are same as \eqref{diffeqn1} and \eqref{diffeqn2}.

\medskip

\subsection{The limit of the Virasoro global block}

In this sub-section we will analyse the non-relativistic limit of the global CFT blocks \eqref{CFF_global_blocks} 
\be
\g^{21}_{34}(p|z,\zb) = z^{h_p-h_1-h_2}\zb^{\hb_p-\hb_1-\hb_2}\hyp(h_p+h_{12},h_p+h_{34};2h_p; z)\hyp(\hb_p+\hb_{12},\hb_p+\hb_{34};2\hb_p;\zb)\nonumber, 
\ee
and check if it matches with the global BMS blocks \eqref{BMS_global_blocks} calculated from the intrinsic analysis.
In order to take the non-relativistic limit of the global block, we make use of the integral representation of the hypergeometric function
\begin{equation}
  \hyp(a,b;c;z) = \frac{\Gamma(c)}{\Gamma(b)\Gamma(c-b)}\int_0^1 dw w^{b-1}(1-w)^{c-b-1}(1-zw)^{-a}
\end{equation}
which is valid for $|z|<1$ with $\arg(z)<\pi$. The beta function also has an integral representation
\be 
\frac{\Gamma(b)\Gamma(c-b)}{\Gamma(c)}=\int_0^1 y^{b-1}(1-y)^{c-b-1} dy.
\ee
So we have 
\be
\hyp(a,b;c;z)=\frac{\int_0^1 w^{b-1}(1-w)^{c-b-1}(1-zw)^{-a}dw}{\int_0^1 y^{b-1}(1-y)^{c-b-1}dy }.
\ee

We consider first the anti-holomorphic sector and make the substitutions $\hb=\frac{\Delta}{2}+\frac{\xi}{2\epsilon}$ and $\zb=t-\epsilon x$ 
\bea
&&\hyp\left(\frac{\Delta_p+\Delta_{12}}{2}+\frac{\xi_p+\xi_{12}}{2\epsilon},\frac{\Delta_p+\Delta_{34}}{2}+\frac{\xi_p+\xi_{34}}{2\epsilon};\Delta_p+\frac{\xi_p}{\epsilon};t-\epsilon x\right)\cr
=&&\frac{\int_0^1 w^{\frac{\Delta_p+\Delta_{34}}{2}+\frac{\xi_p+\xi_{34}}{2\epsilon}-1}(1-w)^{\frac{\Delta_p-\Delta_{34}}{2}+\frac{\xi_p-\xi_{34}}{2\epsilon}-1}\left(1-(t-\epsilon x)\right)^{-\frac{\Delta_p+\Delta_{12}}{2}-\frac{\xi_p+\xi_{12}}{2\epsilon}}dw}{\int_0^1  y^{\frac{\Delta_p+\Delta_{34}}{2}+\frac{\xi_p+\xi_{34}}{2\epsilon}-1}(1-y)^{\frac{\Delta_p-\Delta_{34}}{2}+\frac{\xi_p-\xi_{34}}{2\epsilon}-1}dy}\cr
\equiv && \frac{\bar{I}(\Delta_p,\xi_p,\Delta_{12},\Delta_{34},\xi_{12},\xi_{34})}{\bar{B}(\Delta_p,\xi_p,\Delta_{34},\xi_{34})}.
\eea
Arranging the integrand in $\bar{I}$ as an exponential of a power series in $\epsilon$, we have
\bea
&&\bar{I}(\Delta_p,\xi_p,\Delta_{12},\Delta_{34},\xi_{12},\xi_{34})\cr
&&=\int_0^1 \exp\left(\left(\frac{\Delta_p+\Delta_{34}}{2}+\frac{\xi_p+\xi_{34}}{2\epsilon}-1\right)\log w + \left(\frac{\Delta_p-\Delta_{34}}{2}+\frac{\xi_p-\xi_{34}}{2\epsilon}-1\right)\log(1-w)\right.\cr
  &&\quad\quad\quad\quad\quad\left.- \left(\frac{\Delta_p+\Delta_{12}}{2}+\frac{\xi_p+\xi_{12}}{2\epsilon}\right)\log(1-(t-\epsilon x)w)\right)dw\cr
  &&\equiv \int_0^1 dw \bar{f}(w)e^{\frac{1}{\e}\bar{S}(w)+\O(\epsilon)}, 
\eea
where
\bea
\bar{f}(w)&=& w^{\frac{\Delta _p+\Delta _{34}}{2}-1}(1-w)^{\frac{\Delta _p-\Delta _{34}}{2}-1}(1-tw)^{-\frac{\Delta _p+\Delta _{12}}{2}}e^{\frac{1}{2}\frac{w x \left(\xi _{12}+\xi _p\right)}{-1+t w}},\cr
\bar{S}(w)&=&\frac{1}{2}\left((\xi_p+\xi_{34})\log w + (\xi_p - \xi_{34})\log(1-w) - (\xi_p + \xi_{12})\log(1 - t w)\right). 
\eea
The critical points of the function $\bar{S}(w)$ are given by
\be
\partial_w \bar{S}(w)=0, 
\ee
and are located at
\be
w_{\pm}=\frac{t \xi _{12}-t \xi _{34}-2 \xi _p \pm \sqrt{\left(-t \xi _{12}+t \xi _{34}+2 \xi _p\right){}^2-4 \left(-\xi _{34}-\xi _p\right) \left(t \xi _{12}-t \xi _p\right)}}{2 \left(t \xi _{12}-t \xi _p\right)}. 
\ee
In particular, the critical point $w_-$ lies on the real axis in the domain of integration. 
Then, in the limit $\epsilon\rightarrow0$ limit, we can use the saddle point approximation to calculate the integral
\bea
&&\bar{I}(\Delta_p,\xi_p,\Delta_{12},\Delta_{34},\xi_{12},\xi_{34}) \stackrel{\epsilon\rightarrow0}{\longrightarrow} \sqrt{\frac{2\pi\epsilon}{-\bar{S}''(w_-)}}e^{\frac{1}{\epsilon}\bar{S}(w_-)}\left(\bar{f}(w_-)+\O(\epsilon)\right)\cr
=&&\frac{2\sqrt{\pi \epsilon}  e^{\frac{w_- x \left(\xi _{12}+\xi _p\right)}{-2+2 t w_-}} (1-w_-)^{-1+\frac{-\epsilon  \Delta _{34}+\epsilon  \Delta _p-\xi _{34}+\xi _p}{2 \epsilon }} {w_-}^{\frac{-2 \epsilon +\epsilon  \Delta _{34}+\epsilon  \Delta _p+\xi _{34}+\xi _p}{2 \epsilon }} (1-t w_-)^{-\frac{\epsilon  \Delta _{12}+\epsilon  \Delta _p+\xi _{12}+\xi _p}{2 \epsilon }}}{\sqrt{-\frac{t^2 \left(\xi _{12}+\xi _p\right)}{(-1+t w_-)^2}+\frac{-\xi _{34}+\xi _p}{(-1+w_-)^2}+\frac{\xi _{34}+\xi _p}{{w_-}^2}}}.\cr
&&
\eea
Now, let us do the saddle point analysis for the beta function 
\bea
\bar{B}(\Delta_p,\xi_p,\Delta_{34},\xi_{34})
&=&\int_0^1  y^{\frac{\Delta_p+\Delta_{34}}{2}-1}(1-y)^{\frac{\Delta_p-\Delta_{34}}{2}-1}e^{\frac{1}{2\epsilon}\left((\xi_p+\xi_{34})\log y +(\xi_p+\xi_{34})\log (1-y)\right)} dy\cr
&\equiv &\int_0^1  \bar{p}(y)e^{\frac{1}{\epsilon}\bar{q}(y)} dy,
\eea
where 
\be
\bar{p}(y)=y^{\frac{\Delta_p+\Delta_{34}}{2}-1}(1-y)^{\frac{\Delta_p-\Delta_{34}}{2}-1},\,\,\,\bar{q}(y)=\frac{1}{2}\left((\xi_p+\xi_{34})\log y +(\xi_p+\xi_{34})\log (1-y)\right).
\ee
The critical point for $\bar{q}(y)$ is at $y_{-}=\frac{\xi _{34}+\xi _p}{2 \xi _p}$. Then we have
\bea
\bar{B}(\Delta_p,\xi_p,\Delta_{34},\xi_{34}) &\stackrel{\epsilon\rightarrow0}{\longrightarrow}& \sqrt{\frac{2\pi\epsilon}{-\bar{q}''(y_-)}}e^{\frac{1}{\epsilon}\bar{q}(w_-)} \bar{p}(y_-)\cr
&=& \frac{2 \sqrt{\pi \epsilon }\text{  }\left(1-y_-\right){}^{-\frac{2 \epsilon +\epsilon  \Delta _{34}-\epsilon  \Delta _p+\xi _{34}-\xi _p}{2 \epsilon }} y_-^{\frac{-2 \epsilon +\epsilon  \left(\Delta _{34}+\Delta _p\right)+\xi _{34}+\xi _p}{2 \epsilon }}}{\sqrt{\frac{\left(1-2 y_-\right) \xi _{34}+\left(1+2 \left(-1+y_-\right) y_-\right) \xi _p}{\left(-1+y_-\right){}^2 y_-^2}}}.
\eea
All the arguments above also follows through for the holomorphic sector where we make the substitution $h=\frac{\Delta}{2}-\frac{\xi}{2\epsilon}$ and $z=t+\epsilon x$ 
\bea
&&\hyp\left(\frac{\Delta_p+\Delta_{12}}{2}-\frac{\xi_p+\xi_{12}}{2\epsilon},\frac{\Delta_p+\Delta_{34}}{2}-\frac{\xi_p+\xi_{34}}{2\epsilon};\Delta_p-\frac{\xi_p}{\epsilon};t+\epsilon x\right)\cr
&& = \frac{ \int_0^1 f(w)e^{\frac{1}{\e}S(w)+\O(\epsilon)}dw}{\int_0^1  p(w)e^{\frac{1}{\e}q(w)}dy} \equiv  \frac{I(\Delta_p,\xi_p,\Delta_{12},\Delta_{34},\xi_{12},\xi_{34})}{B(\Delta_p,\xi_p,\Delta_{34},\xi_{34})},
\eea
with
\bea
f(w)&=&w^{\frac{\Delta _p+\Delta _{34}}{2}-1}(1-w)^{\frac{\Delta _p-\Delta _{34}}{2}-1}(1-tw)^{-\frac{\Delta _p+\Delta _{12}}{2}}e^{\frac{1}{2}\frac{w x \left(\xi _{12}+\xi _p\right)}{-1+t w}},\cr
S(w)&=& \frac{1}{2}\left(-(\xi_p+\xi_{34})\log w -(\xi_p - \xi_{34})\log(1-w) + (\xi_p + \xi_{12})\log(1 - t w)\right),\cr
p(y)&=&y^{\frac{\Delta_p+\Delta_{34}}{2}-1}(1-y)^{\frac{\Delta_p-\Delta_{34}}{2}-1},\cr
q(y)&=& \frac{1}{2}\left(-(\xi_p+\xi_{34})\log y -(\xi_p+\xi_{34})\log (1-y)\right).
\eea
It can be easily seen that for the integrand in $I(\Delta_p,\xi_p,\Delta_{12},\Delta_{34},\xi_{12},\xi_{34})$, the dominant saddle occurs at the same value $w_-$. Then the saddle point approximation give us
\bea
&&I(\Delta_p,\xi_p,\Delta_{12},\Delta_{34},\xi_{12},\xi_{34})\cr
=&&\frac{2 \sqrt{\pi \epsilon} e^{\frac{w_- x \left(\xi _{12}+\xi _p\right)}{-2+2 t w_-}} (1-w_-)^{-1+\frac{-\epsilon  \Delta _{34}+\epsilon  \Delta _p+\xi _{34}-\xi _p}{2 \epsilon }} {w_-}^{-\frac{2 \epsilon -\epsilon  \Delta _{34}-\epsilon  \Delta _p+\xi _{34}+\xi _p}{2 \epsilon }} (1-t w_-)^{\frac{-\epsilon  \Delta _{12}-\epsilon  \Delta _p+\xi _{12}+\xi _p}{2 \epsilon }}}{\sqrt{-\frac{\xi _{34}+\xi_p}{{w_-}^2}+\frac{\xi _{34}-\xi _p}{(-1+w_-)^2}+\frac{t^2 \left(\xi _{12}+\xi _p\right)}{(-1+t w_-)^2}}}.\cr
&&
\eea
Similarly for the integrand in $B(\Delta_p,\Delta_{34},\xi_{34})$, the saddle point is same as in the anti-holomorphic sector i.e., $y_{-}$. Doing the saddle point approximation, we have
\be
B(\Delta_p,\Delta_{34},\xi_{34})=\frac{2 i \sqrt{\pi \epsilon }\text{  }\left(1-y_-\right){}^{-\frac{2 \epsilon +\epsilon  \Delta _{34}-\epsilon  \Delta _p-\xi _{34}+\xi _p}{2 \epsilon }} \left(-1+y_-\right) y_-^{1-\frac{2 \epsilon -\epsilon  \Delta _{34}-\epsilon  \Delta _p+\xi _{34}+\xi _p}{2 \epsilon }}}{\sqrt{\left(1-2 y_-\right) \xi _{34}+\left(1-2 y_-+2 y_-^2\right) \xi _p}}. 
\ee 
Now, combining the holomorphic and the anti-holomorphic pieces, we have
\bea
&& \frac{I(\Delta_p,\xi_p,\Delta_{12},\Delta_{34},\xi_{12},\xi_{34})\bar{I}(\Delta_p,\xi_p,\Delta_{12},\Delta_{34},\xi_{12},\xi_{34})}{B(\Delta_p,\Delta_{34},\xi_{34})\bar{B}(\Delta_p,\Delta_{34},\xi_{34})} \cr
=&& i\epsilon \frac{ 4 e^{\frac{w_- x \left(\xi _{12}+\xi _p\right)}{-1+t w_-}} \pi  (1-w_-)^{-2-\Delta _{34}+\Delta _p} {w_-}^{-2+\Delta _{34}+\Delta _p} (1-t w_-)^{-\Delta _{12}-\Delta _p}}
{{\left(\frac{\xi _{34}}{{w_-}^2}+\frac{\xi _p}{{w_-}^2}-\frac{t^2 \left(\xi _{12}+\xi _p\right)}{(-1+t w_-)^2}+\frac{-\xi _{34}+\xi _p}{(-1+w_-)^2}\right)^2}}\cr
&&\times \left( \frac{4 \pi  \epsilon  \left(1-y_-\right){}^{-2-\Delta _{34}+\Delta _p} y_-^{-2+\Delta _{34}+\Delta _p}}{\sqrt{-\frac{\left(\left(1-2 y_-\right) \xi _{34}+\left(1-2 y_-+2 y_-^2\right) \xi _p\right){}^2}{\left(-1+y_-\right){}^4 y_-^4}}}\right)^{-1}.
\eea
We are interested in the case where all the external operators are identical i.e., $\Delta_{ij}=0,\,\xi_{ij}=0$. Putting the value of $w_{-}$ and $y_{-}$ and taking $\Delta_{ij}=0,\,\xi_{ij}=0$ in the above equation, we have
\be
\frac{I(\Delta_p,\xi_p)\bar{I}(\Delta_p,\xi_p)}{B(\Delta_p,\xi_p)\bar{B}(\Delta_p,\xi_p)}= 2^{2\Delta_p-2} \left(1-t\right)^{-1/2}  \exp{\left(\frac{-\xi_p x}{t\sqrt{1-t}} +\xi_p \frac{x}{t} \right)}(1+\sqrt{1-t})^{2-2\D _p}.
\ee
Combining this with the factor
\be
z^{h_p-2h}\bar{z}^{\h_p-2\h} \stackrel{\epsilon\rightarrow0}{\longrightarrow} t^{\Delta_p-2\Delta}e^{-\xi_p \frac{x}{t}+2\xi\frac{x}{t}},
\ee
we finally have the global BMS blocks
\bea{}
g_{\D,\xi}(p|t,x)= 2^{2 \D _p-2}\, \left(1-t\right)^{-1/2}  \exp{\left(\frac{-\xi_p x}{t\sqrt{1-t}} +2\xi \frac{x}{t} \right)} &&t^{\D _p-2\D} (1+\sqrt{1-t})^{2-2\D _p}.\cr
&&
\eea
This matches exactly with global BMS blocks \eqref{BMS_global_blocks} obtained using intrinsic analysis and gives us a very non-trivial and comprehensive check of our intrinsic analysis. 

\newpage

\section{The Chiral Limit}

In this section, we will explore what is called the Chiral limit of the BMS$_3$ algebra. When one is looking at the BMS$_3$ algebra \refb{gca2d} with $c_M=0$ and furthermore, restricting to a sector where $\xi=0$, i.e. a sector where all the $M_0$ eigenvalues of the states considered are vanishing, through an analysis of null vectors in the algebra \cite{Bagchi:2009pe}, it can be shown there is a truncation of the BMS$_3$ down to its Virasoro sub-algebra.  

\subsection{A Holographic Interlude: Flatspace Chiral Gravity}

In the context of holography, the phenomenon of truncation of the symmetry algebra in the field theory has been used to construct what is called Flatspace Chiral Gravity \cite{Bagchi:2012yk}. Topologically Massive Gravity (TMG) is a theory of gravity in 3 dimensions, which, in addition to the usual Einstein Hilbert term, has a gravitational Chern-Simons term. 
\be\label{acttmg}
S_{TMG} = S_{EH} + \frac{1}{\mu} S_{GCS} = \int d^3x \sqrt{-g} \left[R + 2 \Lambda + \frac{1}{\mu} \varepsilon^{\lambda \mu \nu} \Gamma^{\rho}_{\ \lambda \sigma} \left( \p_\mu \Gamma^{\sigma}_{\  \rho \nu} + \frac{2}{3} \Gamma^{\sigma}_{\ \mu \tau} \Gamma^{\tau}_{\ \nu \rho} \right) \right].
\ee
Here $\Lambda$ is the cosmological constant. When one analyses the asymptotic structure of TMG with asymptotically Minkowskian boundary conditions ($\Lambda=0$), the ASG turns out to be the BMS$_3$ algebra, now with both central charges turned on. They take values:
\be
c_L^{tmg} = \frac{1}{4 \mu G}, \quad c_M^{tmg} = \frac{1}{4G}.
\ee
Now if we look at a limit where $G\to \infty, \ \mu\to 0$ with $\mu G = \frac{1}{96k}$ held fixed, the central charges take the value:
\be
c_L = 24 k, \quad c_M=0.
\ee
In this limit, the gravitational C-S term in the TMG action \refb{acttmg} becomes important and the Einstein-Hilbert term is scaled away. This theory is called Chern-Simons Gravity. With asymptotically Minkowskian boundary conditions, this theory has an asymptotic algebra which is just a single copy of a Virasoro algebra with central charge $c=24k$. This theory has been named Flatspace Chiral Gravity (FCG). It can be checked through a gravitational analysis that for FCG, all the $M_n$ charges vanish identically \cite{Bagchi:2012yk}.

\subsection{Chiral limit and the BMS bootstrap}

In this subsection, we wish to see how our earlier analysis of the BMS bootstrap ties in with the chiral limit described above. To this end, we attempt to construct the coefficients of the BMS OPE where we have vanishing $\xi$ weights and $c_M=0$.

For this special limit, $c_M=0,\,\xi=0,\,\xi_p=0$, the recursion relations given by equation \eqref{recurL}, \eqref{recurM}, \eqref{recurM0} reduce to 
\bea
L_{n}|N+n,\a\>_p & = & \left(N+n \a -\D +n \D +\D _p\right)|N,\a\>_p 
\label{recurVL}
\eea 
\be
 M_{n}|N+n,\a\>_p=-(\a+1)|N,\a+1\>_p
\label{recurVM}
\ee
\be
M_{0}|N,\a\>_p	= -(\a+1)|N,\a+1\>_p.
\label{recurVM0}
\ee
Using the above equations, let us try to find the coefficients of the OPE upto 
level two descendants. For level zero we trivially have
\be
|N=0,\a=0\>=\b_{12}^{p\{0,0\},0}|\D_{p},0\>=|\D_{p},0\>.
\ee

\paragraph{Level 1:}
States at level 1 are given by
\be
|1,\a\>_p=\b_{12}^{p\{1,0\},\a}L_{-1}|\D_{p},0\>+\b_{12}^{p\{0,1\},\a}M_{-1}|\D_{p},0\>,\,\,\a=0,1.
\ee
Let us note that for $\xi_p=0$
\bea
M_0|1,\a\>_p &=&  \b_{12}^{p\{1,0\},\a} M_{-1}|\D_{p},\xi_{p}\>,\\
M_1|1,\a\>_p &=& 0, \\
L_1|1,\a\>_p &=& 2\D_p \b_{12}^{p\{1,0\},\a}|\D_{p},\xi_{p}\>.
\eea
Using \eqref{recurVL} we have 
\bea
&& L_{1}|1,0\>_p = \D_{p}|0,0\>_p \implies 2\D_{p}\b_{12}^{p\{1,0\},0}|\D_{p},0\> = \D_{p}|\D_{p},0\> \cr
\implies && \b_{12}^{p\{1,0\},0}=\frac{1}{2}.
\eea
Using the recursion relation \eqref{recurVM0}, we have
\bea
M_{0}|1,1\>_p = 0 \implies \b_{12}^{p\{1,0\},1}M_{-1}|\D_{p},0\> = 0 \implies \b_{12}^{p\{1,0\},1}=0.
\eea
We see that these are exactly the coefficients that we had for a single Virasoro algebra in the first level as was seen in the table \refb{OPE_CFT} in Sec 2.  {\footnote{$\b_{12}^{p\{0,1\},0}$ remains undetermined in this limit and this is to be expected. We also find that one can determine $\b_{12}^{p\{0,1\},1} = -1/2$. But this is a coefficient coming out of a null state $M_{-1}|\D_{p},0\>$ and has to be neglected.}}

\paragraph{Level 2:} We can continue this analysis to any arbitrary level. At level 2 we have
\bea
|2,\a,\D_{p},0\>	&=&	\b_{12}^{p\{1,1\},\a}L_{-1}M_{-1}|\D_{p},0\>+\b_{12}^{p\{2,0\},\a}L_{-1}L_{-1}|\D_{p},0\>+\b_{12}^{p\{(0,1),0\},\a}L_{-2}|\D_{p},0\>     \cr
		&&+\b_{12}^{p\{0,2\},\a}M_{-1}M_{-1}|\D_{p},0\>+\b_{12}^{p\{0,(0,1)\},\a}M_{-2}|\D_{p},0\>,\,\,\,\a=0,1,2.
\eea 
We will focus on determining $\b_{12}^{p\{2,0\},\a}$ and $\b_{12}^{p\{(0,1),0\},\a}$. The detailed analysis is presented in an appendix. Here we mention the answers: 
\be
\b_{12}^{p\{2,0\},0}=\frac{c_L-12\D-4\D_{p}+c_L\D_{p}+8\D_{p}^{2}}{4(c_L-10\D_{p}+2c_L\D_{p}+16\D_{p}^{2})}
\ee 
\be
\b_{12}^{p\{(0,1),0\},0}=\frac{2\D-\D_{p}+4\D\D_{p}+\D_{p}^{2}}{c_L-10\D_{p}+2c_L\D_{p}+16\D_{p}^{2}}.
\ee 
These again are the answers we would have expected from a single Virasoro algebra with central charge $c_L$. This provides a cross-check of the chiral truncation of the BMS$_3$ algebra.

\section{Concluding Remarks}

\subsection{A summary of our results} 
In this paper, we have built on our initial analysis of the BMS bootstrap in \cite{Bagchi:2016geg}. We have provided a lot of detailed calculations that was missing in \cite{Bagchi:2016geg} and also presented a significant amount of new material, the most significant of which is a comprehensive check of all the main results of the intrinsic analysis by a systematic limiting procedure. 

We have first shown, that one could look at the highest weight representations of the BMS$_3$ algebra and, in a manner similar to closely following the conformal bootstrap approach, set up the bootstrap equations in BMS invariant 2d field theories. For this, the central idea was the construction of the BMS operator product expansion, which then allowed us to make statements about the correlation functions of the theory. We made an ansatz for the BMS OPE and showed that this was indeed a consistent choice to make. The OPE could be, e.g., used to check the form of the two and three point functions that were earlier determined from symmetry. 

We then went on to construct recursion relations between primary states under the action of the various modes of the BMS algebra. These recursion relations then allowed us to fix the undetermined coefficients in the OPE completely. We showed the results up to the second level and in principle, this is an analysis that could be continued to any arbitrary order. Using the OPE, we then considered BMS four-point functions and constructed a notion of crossing symmetry for these field theories, which turned out to be different from the usual conformal crossing equations. 

After this, we constructed the BMS blocks and using crossing symmetry, formulated the BMS bootstrap equation. These equations, when solved, will lead to all possible BMS-invariant field theories in two dimensions. The equations are obviously very difficult to solve. As an important step towards the solution of these equations, we looked at the large central charge limit of the BMS blocks. In this limit, the BMS blocks reduced to what we called the global blocks which were the ones containing descendants of only $L_{-1}$ and $M_{-1}$. We then used the Casimirs of BMS to construct two second order differential equations for the global blocks, which we could solve explicitly. As emphasised above, this is a rather important step in the programme of classifying all BMS invariant field theories with the help of the BMS bootstrap. 

Here there is point that we should emphasise. We said that we are interested in BMS invariant field theories as they form putative duals of Minkowski spacetimes. When we consider Einstein gravity in the bulk, we have already state in the beginning that $c_L=0$. Does this mean that our analysis for global blocks would not be valid for Einstein gravity? The answer, interesting, is that it would be. For this, let us look back at Table \refb{level2}. We see that the coefficients that correspond to the ``higher" descendants ($L_{-n} M_{-m} |\D, \xi\>_p$ for $n, m \ge 2$) of the BMS primaries are actually suppressed by $c_M$ alone. This can be checked for higher levels as well. So the global block actually requires only $c_M \to \infty$. Hence, this limit works for the theories putatively dual to Einstein gravity. 

We went on to recover some of our answers, especially the coefficients of the BMS OPE through the non-relativistic limit of the Virasoro algebra. The fact that the answers obtained in the intrinsic and in the limiting method matched was a check of the correctness of both methods. At the end, we looked at a special case where the central charge $c_M=0$ and also all $\xi=0$. This is a limit where the BMS$_3$ algebra is known to reduce to a single copy of the Virasoro algebra. We found that the coefficients of the BMS OPE also reduce to that of a regular chiral CFT in this case. 

\subsection{Future directions}

There are several directions that are being currently pursued and others we hope to work on in the near future. Below we present a list of these.

\medskip

\noindent{\underline{\em Super-BMS bootstrap}}: It is natural to try and generalise our analysis to symmetry algebras with supersymmetry. There exist supersymmetric versions of BMS$_3$ or equivalently, the GCA$_2$. In particular, it is of interest to consider what we call the homogeneous and inhomogeneous Super Galilean Conformal Algebras (SGCAs) \cite{Bagchi:2016yyf}. The homogeneous SGCA is given by 
\bea\label{sgcah}
&& [L_n, L_m] = (n-m) L_{n+m} + \frac{c_L}{12} \, (n^3 -n) \delta_{n+m,0} \nonumber\\
&& [L_n, M_m] = (n-m) M_{n+m} + \frac{c_M}{12} \, (n^3 -n) \delta_{n+m,0} \\
&& [L_n, Q^\a_r] = \Big(\frac{n}{2} - r\Big) Q^\a_{n+r}, \quad \{Q^\a_r, Q^\b_s \} = \delta^{\a\b} \left[M_{r+s} + \frac{c_M}{6} \Big(r^2 - \frac{1}{4}\Big)  \delta_{r+s,0} \right]. \nonumber
\eea
In the above, we have only written the non-zero commutation relations. The above algebra can be obtained by a contraction of the 2D $\mathcal{N}=(1,1)$ superconformal algebra where the fermionic generators are scaled in a similar fashion. This algebra (stripped of the $\a, \b$ indices) has also been obtained as the asymptotic symmetries of 3D $\mathcal{N}=1$ supergravity \cite{Barnich:2014cwa, Barnich:2015sca}. Another version of the Super GCA is the inhomogeneous one: 
\bea\label{sgcai}
&& [L_n, L_m] = (n-m) L_{n+m} + \frac{c_L}{12} (n^3 -n) \delta_{n+m,0}, \nonumber\\
&& [L_n, M_m] = (n-m) M_{n+m} + \frac{c_M}{12} (n^3 -n) \delta_{n+m,0}, \\
&& [L_n, G_r] = \Big(\frac{n}{2} -r\Big) G_{n+r}, \ [L_n, H_r] = \Big(\frac{n}{2} -r\Big) H_{n+r}, \ [M_n, G_r] = \Big(\frac{n}{2} -r\Big) H_{n+r}, \nonumber\\
&& \{ G_r, G_s \} = 2 L_{r+s} + \frac{c_L}{3} \Big(r^2 - \frac{1}{4}\Big)   \delta_{r+s,0}, \ \{ G_r, H_s \} = 2 M_{r+s} + \frac{c_M}{3} \Big(r^2 - \frac{1}{4}\Big)   \delta_{r+s,0}.\nonumber
\eea 
Here again the zero commutators are suppressed. This inhomogeneous SGCA can be obtained from the 2D $\mathcal{N}=(1,1)$ superconformal algebra by a different contraction, a contraction which the fermionic generators are scaled in very different ways \cite{Bagchi:2016yyf}. In the context of supergravity, this leads to a an exotic twisted SUGRA theory \cite{Lodato:2016alv}. 

We are at present attempting to construct the bootstrap programme for both these algebras. One of the crucial steps, as in the bosonic case, is the construction of the OPE for the supersymmetric algebra. It is expected that the analysis would generalise to the supersymmetric case in a natural way. 

\medskip

\noindent{\underline{\em BMS Liouville theory}}: Liouville theory is a very important example of a 2d CFT that admits a semi-classical limit. This semi-classical limit also connects Liouville theory with AdS$_3$ gravity and $SL(2,R)$ Chern-Simons theory \cite{Achucarro:1987vz, Witten:1988hc, Coussaert:1995zp}. The three point function for general momenta in Liouville theory has been computed and goes under the name of the DOZZ formula \cite{Dorn:1994xn,Zamolodchikov:1995aa}. Closed form expression for the structure constants are thus known and it has been explicitly verified that the theory satisfies the conformal bootstrap equation \cite{Zamolodchikov:1995aa, Teschner:2001rv}. A lot of the progress on Virasoro blocks in general 2d CFTs hinges on the success of computations in Liouville theory. In the recent emergence of techniques of 2d CFTs for large central charge following \cite{Hartman:2013mia}, Liouville theory has played a central role. 

In \cite{Barnich:2012rz}, a contracted version of Liouville theory has been proposed by taking a systematic limit of the parent theory. The Poisson algebra of the conserved charges of this theory turns out to be the BMS algebra \refb{gca2d}. There are actually two versions of this limiting theory, one with a vanishing $c_M$ and one with $c_M$ non-zero. It is expected that the semi-classical version of the theory with $c_M \neq 0$ would be important for understanding the dual of Einstein gravity in 3d flat spacetimes. 

We wish to understand the structure of this theory in detail so that we can compute the equivalent of the DOZZ formula for the three point functions and hence find an explicit example where the BMS bootstrap equations are satisfied. For our explorations of flat holography, this is a vital step. We wish to address questions about BMS blocks and also generalise the recent large $c$ CFT techniques to the BMS case. The BMS Liouville theory would provide us valuable insight into these problems. 

\medskip

\noindent{\underline{\em Higher dimensions}}: A very natural direction of generalisation of our analysis is to explore a higher dimensional version of our bootstrap analysis. We have made some remarks about this in the introduction. Let us briefly elaborate on some aspects and some possible difficulties. First thing to mention here is that the BMS algebra and the GCA are not isomorphic in higher dimensions, as can be readily seen by looking at equations \refb{bms4} and \refb{GCA}. So higher dimensional generalisations of the bootstrap would be different for the two cases. 

The structure of BMS algebras in different dimensions is very different. This should be obvious by looking at the 4d case \refb{bms4} and the 3d case \refb{gca2d}, which we have addressed in this paper. The systematics of the bootstrap procedure, which depends crucially on the structure of the algebra, would thus be very different and our methods in this paper would not generalise in any natural way for the cases of field theories in 3 and higher dimensions with BMS symmetry. There are indications \cite{Bagchi:2016bcd} that 3d field theories with BMS$_4$ symmetry actually reduce to 2d CFTs\footnote{See also \cite{Kapec:2016jld, Cheung:2016iub}.}. So, it is possible that the usual 2d Virasoro bootstrap would be applicable for these field theories. We don't have any concrete suggestions for field theories with BMS symmetries in even higher dimensions ($D>4$). 

The case for GCFTs and the Galilean Conformal Bootstrap in higher dimension is much more encouraging from the point of view of our present analysis. The structure of Galilean conformal symmetry remains very similar as we go up in dimensions. This is evident from \refb{gca2d} and \refb{GCA}{\footnote{There could have been additional complications from an infinite lift of the rotation generators in higher dimensions. But, interestingly, in the field theories that admit the GCA in higher dimensions like Galilean Electrodynamics and Galilean Yang-Mills theories, the symmetry algebra of relevance in the one where rotations don't get any lift.}}. It is hence expected that constructions similar to what we have attempted in this paper would also work for higher dimensional GCFTs. There would be interesting departures as well. There is no central charge in the $[L_n, M_m^i]$ commutator and hence the semi-classical limit should work differently. 

The very interesting thing in this analysis would be the fact that unlike conformal bootstrap in dimensions higher 2, the Galilean conformal bootstrap would benefit from the infinite dimensional symmetry algebra in all dimensions. This would means we would have much more analytical control over our analysis, as compared to the bootstrap programme in the higher dimensional CFT, which has been driven primary with numerical methods. We should be in a very good position to classify non-relativistic quantum field theories by the virtue of this analysis. 

\medskip

\noindent{\underline{\em Mellin space}}: Recently, it has been shown that one can combine Polykov's original idea about the bootstrap exploiting manifest crossing symmetry with the Mellin representations of CFT amplitudes to get a much better analytical handle on the conformal bootstrap programme for higher dimensions \cite{Gopakumar:2016wkt, Gopakumar:2016cpb}. It is very tempting to attempt a similar algorithm for the BMS bootstrap. Here we would need modifications of the Mellin space amplitudes, for which the systematic limit from CFT should be very useful. 

\bigskip

\noindent There are indeed many other interesting directions to pursue over and above the ones just mentioned. To conclude, we believe the BMS programme that we have initiated in \cite{Bagchi:2016geg} and elaborated on in this paper is a programme which would be very useful in many diverse fields.

\subsection*{Acknowledgements}
This paper is an invited article for the {\em{Classical and Quantum Gravity}} Focus issue on ``BMS Asymptotic Symmetries" edited by Geoffrey Compere. AB thanks Geoffrey for the invitation and for the initiative for editing this special issue. 

\medskip

\noindent It is a pleasure to thank Ivano Lodato, Wout Merbis, Hernan Gonzalez, Glenn Barnich, P. Raman and N. V. Suryanarayana for discussions. AB also wishes to thank the Simons Center for Geometry and Physics for warm hospitality and support during the time of this project. AB is partially supported by an INSPIRE grant from DST, India and by a Max-Planck mobility grant. MG is supported by the FWF project P27396-N27 and the OeAD project IN 03/2015. Z is supported by the India-Israel joint research project UGC/PHY/2014236 and SERB National Post Doctoral Fellowship PDF/2016/002166.

\appendix
\section*{Appendices}
\bigskip
\section{Level 2 coefficients: Detailed calculations in intrinsic method}

In this appendix, we provide further detailed calculations of the analysis outline in Sec 3.4 for finding the coefficients of the OPE for the BMS$_3$ algebra. Below are the details for the level 2 calculations. 

\bigskip

\noindent
At level 2 we have
\bea
|2,\a\>_p & = & \b_{12}^{p\{1,1\},\a}L_{-1}M_{-1}|\D_{p},\xi_{p}\>+\b_{12}^{p\{2,0\},\a}L_{-1}L_{-1}|\D_{p},\xi_{p}\>+\b_{12}^{p\{(0,1),0\},\a}L_{-2}|\D_{p},\xi_{p}\>\cr
 &  & +\b_{12}^{p\{0,2\},\a}M_{-1}M_{-1}|\D_{p},\xi_{p}\>+\b_{12}^{p\{0,(0,1)\},\a}M_{-2}|\D_{p},\xi_{p}\>,\,\,\,\a=0,1,2.
\eea
Let us note again that
\bea
M_0|2,\a\>_p & = & \left(\xi_p\b_{12}^{p\{1,1\},\a}+2\b_{12}^{p\{2,0\},\a}\right)L_{-1}M_{-1}|\D_{p},\xi_{p}\>+\xi_p\b_{12}^{p\{2,0\},\a}L_{-1}L_{-1}|\D_{p},\xi_{p}\>\cr
&&+\xi_p\b_{12}^{p\{(0,1),0\},\a}L_{-2}|\D_{p},\xi_{p}\>+\left(\b_{12}^{p\{1,1\},\a}+\xi_p\b_{12}^{p\{0,2\},\a}\right)M_{-1}M_{-1}|\D_{p},\xi_{p}\>\cr
&&+\left(2\b_{12}^{p\{(0,1),0\},\a}+\xi_p\b_{12}^{p\{0,(0,1)\},\a}\right)M_{-2}|\D_{p},\xi_{p}\>,\\
M_1|2,\a\>_p &=& \left(2\xi_p\b_{12}^{p\{1,1\},\a}+2\b_{12}^{p\{2,0\},\a}+3\b_{12}^{p\{(0,1),0\},\a}\right)M_{-1}|\D_{p},\xi_{p}\>\cr
&&+4\xi_p \b_{12}^{p\{2,0\},\a} L_{-1}|\D_{p},\xi_{p}\>,\\
M_2 |2,\a\>_p &=& \left(6\xi_p \b_{12}^{p\{2,0\},\a} +\left(4\xi_p+\frac{c_M}{2}\right)\b_{12}^{p\{(0,1),0\},\a}\right)|\D_{p},\xi_{p}\>, \\
L_1|2,\a\>_p &=& \left(2(\D_p+1)\b_{12}^{p\{1,1\},\a}+4\xi_p\b_{12}^{p\{0,2\},\a}+3\b_{12}^{p\{0,(0,1)\},\a}\right)M_{-1}|\D_{p},\xi_{p}\>\cr
&&+\left(2\xi_p \b_{12}^{p\{1,1\},\a}+2(2\D_p+1)\b_{12}^{p\{2,0\},\a}+3\b_{12}^{p\{(0,1),0\},\a}\right) L_{-1}|\D_{p},\xi_{p}\>,\\
L_2 |2,\a\>_p &=& \left(6\xi_p \b_{12}^{p\{1,1\},\a}+6\D_p \b_{12}^{p\{2,0\},\a}+\left(4\D_p+\frac{c_L}{2}\right) \b_{12}^{p\{(0,1),0\},\a} \right. \cr
&&+\left. \left(4\xi_p+\frac{c_M}{2}\right)\b_{12}^{p\{0,(0,1)\},\a}\right)|\D_{p},\xi_{p}\>. 
\eea
Using the recursion relation \eqref{recurM0} on the state $|2,2\>_p$, we get
\bea
&&M_{0}|2,2\>_p =  \xi_{p}|2,2\>_p\cr
\implies &&\left(2\b_{12}^{p\{2,0\},2}L_{-1}M_{-1}+\b_{12}^{p\{1,1\},2}M_{-1}M_{-1}+2\b_{12}^{p\{(0,1),0\},2}M_{-2}\right)|\D_{p},\xi_{p}\>  =  0,\cr
\implies && \b_{12}^{p\{1,1\},2}=\b_{12}^{p\{2,0\},2}=\b_{12}^{p\{(0,1),0\},2}=0.
\eea
Using the above result and \eqref{recurL} with $N=0,n=2,\a=2$, we have
\bea
L_{2}|2,2\>_p  =  0 \Rightarrow \left(4\xi_p+\frac{c_M}{2}\right)\b_{12}^{p\{0,(0,1)\},2}|\D_{p},\xi_{p}\>  =  0
\Rightarrow
\b_{12}^{p\{0,(0,1)\},2}=0.
\eea
Using \eqref{recurL} with $N=1,n=1,\a=2$, we have
\bea
&& L_{1}|2,2\>_p  =  -\xi_{p}|1,1\>_p\cr
\implies && 4\xi_p\b_{12}^{p\{0,2\},2}M_{-1}|\D_{p},\xi_{p}\>  =  \frac{\xi_{p}}{2}M_{-1}|\D_{p},\xi_{p}\>\cr
\implies && \b_{12}^{p\{0,2\},2}=\frac{1}{8}.
\eea
Next, we use \eqref{recurM0} on the state $|2,1\>_p$, giving us
\bea
&& M_{0}|2,1\>_p  =  \xi_{p}\,|2,1\>_p-2|2,2\>_p\cr
\implies && \left(2\b_{12}^{p\{2,0\},1}L_{-1}M_{-1}
+\b_{12}^{p\{1,1\},1}M_{-1}M_{-1}+2\b_{12}^{p\{(0,1),0\},1}M_{-2}\right)|\D_{p},\xi_{p}\> \cr
&& = -2\b_{12}^{p\{0,2\},2}M_{-1}M_{-1}|\D_{p},\xi_{p}\> \cr
\implies && 
\b_{12}^{p\{1,1\},1}=-2\b_{12}^{p\{0,2\},2}=-\frac{1}{4},\,\,\b_{12}^{p\{2,0\},1}=0,\,\,\,\b_{12}^{p\{(0,1),0\},1}=0.
\eea
Now, we use \eqref{recurL} with $N=0,n=2,\a=1$
\bea
&& L_{2}|2,1\>_p  =  \left(-2\xi-2\xi_{p}\right)|0,0\>_p\cr
\implies && \left(-\frac{1}{4}6\xi_{p}+\b_{12}^{p\{0,(0,1)\},1}\left(4\xi_{p}+\frac{c_M}{2}\right)\right)|\D_{p},\xi_{p}\>  =  (-2\xi-2\xi_{p})|\D_{p},\xi_{p}\>\cr
\implies && \b_{12}^{p\{0,(0,1)\},1}=-\frac{2\xi+\frac{1}{2}\xi_{p}}{(4\xi_{p}+6c_M)}=-\frac{4\xi+\xi_{p}}{4\left(3c_M+2\xi_{p}\right)}.
 \eea
Using the recursion relation \eqref{recurL} with $N=1,n=1,\a=1$, we have
\bea
&&L_{1}|2,1\>_p  =  (2+\D_{p})|1,1\>_p-\xi_{p}|1,0\>_p\cr
\implies &&\left(-\frac{\xi_p}{2}L_{-1}+\left(-\frac{(1+\D_{p})}{2} + 4\xi_p\b_{12}^{p\{0,2\},1}+3\b_{12}^{p\{0,(0,1)\},1}\right)M_{-1}\right)|\D_{p},\xi_{p}\> \cr
&&= \left(-\frac{(2+\D_{p})}{2}M_{-1}-\frac{\xi_{p}}{2}L_{-1}\right)|\D_{p},\xi_{p}\>\cr
\implies &&
\b_{12}^{p\{0,2\},1}=-\frac{1}{8\xi_{p}}-\frac{3}{4\xi_{p}}\b_{12}^{p\{0,(0,1)\},1}=\frac{12 \xi -6 c_M-\xi _p}{16 \xi _p \left(3 c_M+2 \xi _p\right)}.
\eea
Now, we use \eqref{recurM0} on the state $|2,0\>_p $, giving us
\bea
&& M_{0}|2,0\>_p  =  \xi_{p}\,|2,0\>_p-|2,1\>_p\cr
\implies && \left(\b_{12}^{p\{2,0\},0}2L_{-1}M_{-1} +\b_{12}^{p\{1,1\},0}M_{-1}M_{-1} + 2\b_{12}^{p\{(0,1),0\},0}M_{-2}\right)|\D_{p},\xi_{p}\> \cr
& &= -\left(\b_{12}^{p\{1,1\},1}L_{-1}M_{-1}+\b_{12}^{p\{0,2\},1}M_{-1}M_{-1}+\b_{12}^{p\{0,(0,1)\},1}M_{-2}\right)|\D_{p},\xi_{p}\>\cr
\implies &&\b_{12}^{p\{1,1\},0}=-\b_{12}^{p\{0,2\},1}=-\frac{12 \xi -6 c_M-\xi _p}{16 \xi _p \left(3 c_M+2 \xi _p\right)},\,\,\,\b_{12}^{p\{2,0\},0}=-\frac{\b_{12}^{p\{1,1\},1}}{2}=\frac{1}{8},\cr
&&\b_{12}^{p\{(0,1),0\},0}=-\frac{\b_{12}^{p\{0,(0,1)\},1}}{2}=\frac{4\xi+\xi_{p}}{8\left(3c_M+2\xi_{p}\right)}.
\eea
Next, we use \eqref{recurL} with $n=2,N=0,\a=0$
\bea
&&L_{2}|2,0\>_p=(\D+\D_{p})|0,0\>_p\cr
\implies && \left(6\b_{12}^{p\{1,1\},0}\xi_{p}+6\b_{12}^{p\{2,0\},0}\D_{p}+\b_{12}^{p\{(0,1),0\},0}\left(4\D_{p}+\frac{c_L}{2}\right)+\b_{12}^{p\{0,(0,1)\},0}\left(4\xi_{p}+\frac{c_M}{2}\right)\right)|\D_{p},\xi_{p}\>\cr
&&=(\D+\D_{p})|\D_{p},\xi_{p}\> \cr
\implies && \b_{12}^{p\{0,(0,1)\},0} = \frac{\D+\D_{p}-6\b_{12}^{p\{1,1\},0}\xi_{p}-6\b_{12}^{p\{2,0\},0}\D_{p}-\b_{12}^{p\{(0,1),0\},0}(4\D_{p}+6c_L)}{(4\xi_{p}+6c_M)}\cr
&&=\frac{36 \xi -24 \xi  c_L-18 c_M+24 \D  c_M-16 \xi  \D _p+6 c_M \D _p-3 \xi _p+16 \D  \xi _p-6 c_L \xi _p}{16 \left(3 c_M+2 \xi _p\right)^2}.\cr
&&
\eea
Furthermore \eqref{recurL} with $n=1,N=1,\a=0$ give us the recursion relation

\bea
&& L_{1}|2,0\>_p  =  (1+\D_{p})|1,0\>_p\cr
\implies && \left(2(\D_p+1)\b_{12}^{p\{1,1\},0}+4\xi_p\b_{12}^{p\{0,2\},0}+3\b_{12}^{p\{0,(0,1)\},0}\right)M_{-1}|\D_{p},\xi_{p}\>\cr
&&+\left(2\xi_p \b_{12}^{p\{1,1\},0}+2(2\D_p+1)\b_{12}^{p\{2,0\},0}+3\b_{12}^{p\{(0,1),0\},0}\right) L_{-1}|\D_{p},\xi_{p}\>\cr
&&=\frac{(1+\D_{p})}{2}L_{-1}|\D_{p},\xi_{p}\>
\eea
Equating coefficient of $M_{-1}|\D_{p},\xi_{p}\>$ in the above equation, we get
\bea
\b_{12}^{p\{0,2\},0} & = & -\frac{1}{4\xi_{p}}\left(2\b_{12}^{p\{1,1\},0}\left(\D_{p}+1\right)+3\b_{12}^{p\{0,(0,1)\},0}\right)\cr
&=&\frac{1}{64 \xi _p^2 \left(3 c_M+2 \xi _p\right)^2}\left(-36 c_M^2 \left(1+\D _p\right)+24 c_M \left(3 \xi +\D _p \left(3 \xi -2 \xi _p\right)+(1-3 \D ) \xi _p\right)\right.\cr
&&\left. +\xi _p \left(-60 \xi +\D _p \left(96 \xi -4 \xi _p\right)+5 \xi _p-48 \D  \xi _p+18 c_L \left(4 \xi +\xi _p\right)\right)\right).
\eea
The coefficients are collected in Table 2 in Sec 3.4. 

\bigskip

It is clear from the analysis above that, given the recursion relations, we can solve for the $\beta$s for any level. Computational power required obviously increases substantially as we attempt to go higher, but there is no theoretical difficulty in obtaining these coefficients. 

\newpage

\section{Level 2 coefficients: Detailed calculations in limiting method}
Using the relation
\be
c=\frac{1}{6}\left(c_L-\frac{c_M}{\e}\right),\,\,\,\bar{c}=\frac{1}{6}\left(c_L+\frac{c_M}{\e}\right),\,\,\,h=\frac{1}{2}\left(\D-\frac{\xi}{\e}\right),\,\,\,\h=\frac{1}{2}\left(\D+\frac{\xi}{\e}\right),
\label{hw}
\ee
the coefficients of the level two descendant fields given in Table  \eqref{OPE_CFT} are
\bea
\mathcal{B}_{12}^{p\{1,1\}} & = & \frac{c-12h-4h_{p}+ch_{p}+8h_{p}^{2}}{4(c-10h_{p}+2ch_{p}+16h_{p}^{2})}=\frac{1}{8}+\b\e+\lambda\e^2+\mathcal{O}(\e^{3}),\\
\mathcal{B}_{12}^{p\{2\}} & = & \frac{2h-h_{p}+4hh_{p}+h_{p}^{2}}{c-10h_{p}+2ch_{p}+16h_{p}^{2}}=\eta + \Gamma\e+\mathcal{O}(\e^{2}),\\
\bar{\mathcal{B}}_{12}^{p\{\bar{1},\bar{1}\}} & = & \frac{\bar{c}-12\h-4\h_{p}+\bar{c}\h_{p}+8\h_{p}^{2}}{4(\bar{c}-10\h_{p}+2c\bar{c}\h+16\h_{p}^{2})}=\frac{1}{8}-\b\e+\lambda\e^{2}+\mathcal{O}(\e^{3}),\\
\bar{\mathcal{B}}_{12}^{p\{\bar{2}\}} & = & \frac{2\h-\h_{p}+4\h\h_{p}+\h_{p}^{2}}{c-10\h_{p}+2c\h_{p}+16\h_{p}^{2}}=\eta-\Gamma\e+\mathcal{O}(\e^{2}),
\eea
where 
\bea
\b&=&\frac{12 \xi -6 c_M-\xi _p}{16 \xi _p \left(3 c_M+2 \xi _p\right)}=\b_{12}^{p\{0,2\},1},\\
\lambda&=&\frac{1}{32 \xi _p^2 \left(3 c_M+2 \xi _p\right){}^2}(-36 c_M^2 \left(1+\D _p\right)+24 c_M \left(3 \xi +\D _p \left(3 \xi -2 \xi _p\right)+(1-3 \D ) \xi _p\right)\cr
&&+\xi _p \left(-60 \xi +\D _p \left(96 \xi -4 \xi _p\right)+5 \xi _p-48 \D  \xi _p+18 c_L \left(4 \xi +\xi _p\right)\right))\cr
&=& 2\b_{12}^{p\{0,2\},0},\\
\eta&=&\frac{\left(4\xi+\xi_{p}\right)}{8\left(3c_M+2\xi_{p}\right)}=-\frac{\b_{12}^{p\{0,(0,1)\},1}}{2},\\
\Gamma&=&\frac{\left(-36 \xi +24 \xi  c_L+18 c_M-24 \D  c_M+16 \xi  \D _p-6 c_M \D _p+3 \xi _p-16 \D  \xi _p+6 c_L \xi _p\right)}{16 \left(3 c_M+2 \xi _p\right){}^2}\cr
&=&-\b_{12}^{p\{0,(0,1)\},0}.
\eea

Now collecting all the level two states in the expansion \eqref{ope_nrlimit}, modulo the common factor $\frak{C}_{p12}t^{\D-2\D_{p}}\exp\left(2\xi-\xi_{p})\right)$, we get
\bea
 &  &\left(\mathcal{B}_{12}^{p\{2\}}t^{2}\left(1+\e\frac{x}{t}\right)^{2}\L_{-2}+\mathcal{B}_{12}^{p\{1,1\}}t^{2}\left(1+\e\frac{x}{t}\right)^{2}\L_{-1}\L_{-1}\right.\cr
 && \left. +\mathcal{B}_{12}^{p\{1\}}\bar{\mathcal{B}}_{12}^{p\{\bar{1}\}}t^{2}\left(1+\e\frac{x}{t}\right)\left(1-\e\frac{x}{t}\right)\L_{-1}\mathcal{\bar{L}}_{-1} \right.\cr
 &  & \left. +\bar{\mathcal{B}}_{12}^{p\{\bar{2}\}}t^{2}(1-\e\frac{x}{t})^{2}\mathcal{\bar{L}}_{-2}+\bar{\mathcal{B}}_{12}^{p\{\bar{1},\bar{1}\}}t^{2}(1-\e\frac{x}{t})^{2}\mathcal{\bar{L}}_{-1}\mathcal{\bar{L}}_{-1}\right)|\D_{p},\xi_{p}\>\cr
 & = & \left(\mathcal{B}_{12}^{p\{2\}}t^{2}(1+\e\frac{x}{t})^{2}\frac{1}{2}(L_{-2}-\frac{{1}}{\e}M_{-2})+\mathcal{B}_{12}^{p\{1,1\}}t^{2}(1+\e\frac{x}{t})^{2}\frac{1}{2}(L_{-1}-\frac{{1}}{\e}M_{-1})\frac{1}{2}(L_{-1}-\frac{{1}}{\e}M_{-1}) \right. \cr
 &  & \left. +\mathcal{B}_{12}^{p\{1\}}\bar{\mathcal{B}}_{12}^{p\{\bar{1}\}}t^{2}(1+\e\frac{x}{t})(1-\e\frac{x}{t})\frac{1}{2}(L_{-1}-\frac{{1}}{\e}M_{-1})\frac{1}{2}(L_{-1}+\frac{_{1}}{\e}M_{-1}) \right. \cr
 &  & \left. +\bar{\mathcal{B}}_{12}^{p\{\bar{2}\}}t^{2}(1-\e\frac{x}{t})^{2}\frac{1}{2}(L_{-2}+\frac{{1}}{\e}M_{-2})+\bar{\mathcal{B}}_{12}^{p\{\bar{1},\bar{1}\}}t^{2}(1-\e\frac{x}{t})^{2}\frac{1}{2}(L_{-1} \right. \cr
&&\left. +\frac{{1}}{\e}M_{-1})\frac{1}{2}(L_{-1}+\frac{{1}}{\e}M_{-1}) \right)|\D_{p},\xi_{p}\>.
\label{level2expnasion}
\eea
Then the non-relativistic limit of the coefficient of $x^2$ in the above state is given by
\bea
|2,2\>_p &=&\lim_{\e\longrightarrow 0}\left(\frac{\mathcal{B}_{12}^{p\{2\}}}{2}(\e^{2}L_{-2}-\e M_{-2})+\frac{\mathcal{B}_{12}^{p\{1,1\}}}{4}(\e^{2}L_{-1}L_{-1}-2\e L_{-1}M_{-1}+M_{-1}M_{-1}) \right. \cr
 &  & \left. -\frac{\mathcal{B}_{12}^{p\{1\}}\bar{\mathcal{B}}_{12}^{p\{\bar{1}\}}}{4}(\e^{2}L_{-1}L_{-1}-M_{-1}M_{-1}) -\frac{\bar{\mathcal{B}}_{12}^{p\{\bar{2}\}}}{2}(\e^{2}L_{-2}+\e M_{-2}) \right. \cr
 &  &\left. +\frac{\bar{\mathcal{B}}_{12}^{p\{\bar{1},\bar{1}\}}}{4}(\e^{2}L_{-1}L_{-1}+2\e L_{-1}M_{-1}+M_{-1}M_{-1})\right)|\D_p,\xi_p\> \cr
 &=&\frac{1}{8}M_{-1}M_{-1}|\D_p,\xi_p\>=\b_{12}^{p\{0,2\},2}M_{-1}M_{-1}|\D_p,\xi_p\>.
\eea
Similarly, the coefficients of $xy$ in \eqref{level2expnasion} gives the state 
\bea
|2,1\>_p &=& \lim_{\e\to0}\left(\mathcal{B}_{12}^{p\{2\}}(\e L_{-2}-M_{-2})+\frac{\mathcal{B}_{12}^{p\{1,1\}}}{2}(\e L_{-1}L_{-1}-2L_{-1}M_{-1}+\frac{1}{\e}M_{-1}M_{-1})\right.
\cr
&&\left. -\bar{\mathcal{B}}_{12}^{p\{\bar{2}\}}(\e L_{-2}+M_{-2})-\frac{\bar{\mathcal{B}}_{12}^{p\{\bar{1},\bar{1}\}}}{2}(\e L_{-1}L_{-1}+2L_{-1}M_{-1}+\frac{1}{\e}M_{-1}M_{-1})\right)|\D_{p},\xi_{p}\>\cr
&=&\left(-\frac{1}{4}L_{-1}M_{-1}-2\eta M_{-2}+\b M_{-1}M_{-1}\right)|\D_{p},\xi_{p}\>\cr
&=&\left(\b_{12}^{p\{1,1\},1}L_{-1}M_{-1}+\b_{12}^{p\{0,(0,1)\},1}M_{-2}+\b_{12}^{p\{0,2\},1}M_{-1}M_{-1}\right)|\D_{p},\xi_{p}\>.
\eea
Lastly, the non-relativistic limit of the coefficients of $t^2$ gives us the state 
\bea
|2,0\>_p	&=&	\lim_{\e\to0}\left(\frac{\mathcal{B}_{12}^{p\{2\}}}{2}(L_{-2}-\frac{1}{\e}M_{-2})+\frac{\mathcal{B}_{12}^{p\{1,1\}}}{4}(L_{-1}L_{-1}-\frac{2}{\e}L_{-1}M_{-1}+\frac{1}{\e^{2}}M_{-1}M_{-1}) \right. \cr
		&&\left. +\frac{\mathcal{B}_{12}^{p\{1\}}\bar{\mathcal{B}}_{12}^{p\{\bar{1}\}}}{4}(L_{-1}L_{-1}-\frac{1}{\e^{2}}M_{-1}M_{-1})+\frac{\bar{\mathcal{B}}_{12}^{p\{\bar{2}\}}}{2}(L_{-2}+\frac{{1}}{\e}M_{-2}) \right. \cr
		&& \left. +\frac{\bar{\mathcal{B}}_{12}^{p\{\bar{1},\bar{1}\}}}{4}(L_{-1}L_{-1}+\frac{2}{\e}L_{-1}M_{-1}+\frac{1}{\e^{2}}M_{-1}M_{-1})\right)|\D_{p},\xi_{p}\>\cr
		&=&\left(\eta L_{-2}-\Gamma M_{-2}+\frac{1}{8}L_{-1}L_{-1}-\b L_{-1}M_{-1}+\frac{\lambda}{2}M_{-1}M_{-1}\right)|\D_{p},\xi_{p}\>\cr
&=&\left(\b_{12}^{p\{(0,1),0\},0} L_{-2}+\b_{12}^{p\{0,(0,1)\},0} M_{-2}+\b_{12}^{p\{2,0\},0}L_{-1}L_{-1}+\b_{12}^{p\{1,1\},0}L_{-1}M_{-1} \right. \cr
&&\left. +\b_{12}^{p\{0,2\},0}M_{-1}M_{-1}\right)|\D_{p},\xi_{p}\>.		
 \eea
All of these states match with our calculations in the previous section.

\newpage

\section{Level 2 analysis of coefficients in the Chiral limit}
\bea
|2,\a,\D_{p},0\>	&=&	\b_{12}^{p\{1,1\},\a}L_{-1}M_{-1}|\D_{p},0\>+\b_{12}^{p\{2,0\},\a}L_{-1}L_{-1}|\D_{p},0\>+\b_{12}^{p\{(0,1),0\},\a}L_{-2}|\D_{p},0\>     \cr
		&&+\b_{12}^{p\{0,2\},\a}M_{-1}M_{-1}|\D_{p},0\>+\b_{12}^{p\{0,(0,1)\},\a}M_{-2}|\D_{p},0\>,\,\,\,\a=0,1,2.
\eea 
Note that for $\xi_p=0,\,c_M=0$. 
\bea
M_0|2,\a\>_p & = & 2\b_{12}^{p\{2,0\},\a}L_{-1}M_{-1}|\D_{p},0\> + \b_{12}^{p\{1,1\},\a}M_{-1}M_{-1}|\D_{p},0\>\cr
&&+2\b_{12}^{p\{(0,1),0\},\a}M_{-2}|\D_{p},0\>,\\
M_1|2,\a\>_p &=& (2\b_{12}^{p\{2,0\},\a}+3\b_{12}^{p\{(0,1),0\},\a})M_{-1}|\D_{p},0\>\\
M_2 |2,\a\>_p &=& 0, \\
L_1|2,\a\>_p &=& (2(\D_p+1)\b_{12}^{p\{1,1\},\a}+3\b_{12}^{p\{0,(0,1)\},\a})M_{-1}|\D_{p},0\>\cr
&&+(2(2\D_p+1)\b_{12}^{p\{2,0\},\a}+3\b_{12}^{p\{(0,1),0\},\a}) L_{-1}|\D_{p},0\>,\\
L_2 |2,\a\>_p &=& \left(6\D_p \b_{12}^{p\{2,0\},\a}+\left(4\D_p+ \frac{c_L}{2}\right) \b_{12}^{p\{(0,1),0\},\a}\right)|\D_{p},0\>. 
\eea
Using the recursion relation \eqref{recurVM0}, we have
\bea  
&& M_{0}|2,2\>_p	=	0 \cr
\implies && 2\b_{12}^{p\{2,0\},\a}L_{-1}M_{-1}|\D_{p},0\> + \b_{12}^{p\{1,1\},\a}M_{-1}M_{-1}|\D_{p},0\>	=	0 \cr
\implies && \b_{12}^{p\{1,1\},2}=\b_{12}^{p\{2,0\},2}=\b_{12}^{p\{(0,1),0\},2}=0.
\eea 
Now using \eqref{recurVL} we have
\bea 
&& L_{1}|2,2\>_p=0
\implies  \b_{12}^{p\{0,(0,1)\},2}	=	0.
\eea
We also have
 \bea
&& M_{0}|2,1\>_p	=	-2|2,2\>_p \cr
\implies && \left(2\b_{12}^{p\{2,0\},1}L_{-1}M_{-1} + \b_{12}^{p\{1,1\},1}M_{-1}M_{-1}+2\b_{12}^{p\{(0,1),0\},\a}M_{-2}\right)|\D_{p},0\>\cr
&&  =	-2\b_{12}^{p\{0,2\},2}M_{-1}M_{-1}|\D_{p},\>,\cr
\implies && \b_{12}^{p\{1,1\},1}=-2\b_{12}^{p\{0,2\},2},\,\,\b_{12}^{p\{2,0\},1}=0,\,\,\,\b_{12}^{p\{(0,1),0\},1}=0,
 \eea
\bea
&& L_{1}|2,1\>_p	=	(4+\D_{p})|1,1\>_p \cr
\implies && 2(1+\D_{p})\b_{12}^{p\{1,1\},1}+3\b_{12}^{p\{0,(0,1)\},1}	=	\frac{1}{2}(4+\D_{p}) .
\eea
Furthermore
\bea
&& M_{0}|2,0\>_p	=	-|2,1\>_p \cr
\implies && \left(\b_{12}^{p\{1,1\},0}M_{-1}M_{-1}+\b_{12}^{p\{2,0\},0}2L_{-1}M_{-1}+2\b_{12}^{p\{(0,1),0\},0}M_{-2}\right)|\D_{p},0\> \cr
&&=	-\left(\b_{12}^{p\{1,1\},1}L_{-1}M_{-1}+\b_{12}^{p\{0,2\},1}M_{-1}M_{-1}+\b_{12}^{p\{0,(0,1)\},1}M_{-2}\right)|\D_{p},0\> \cr
\implies &&\b_{12}^{p\{1,1\},0}=-\b_{12}^{p\{0,2\},1},\,\,\,2\b_{12}^{p\{2,0\},0}=-\b_{12}^{p\{1,1\},1},\,\,\,2\b_{12}^{p\{(0,1),0\},0}=-\b_{12}^{p\{0,(0,1)\},1}.\cr
&&
\eea 
Using \eqref{recurVL} we have
\bea
&& L_{1}|2,0\>_p	=	(1+\D_{p})|1,0\>_p \cr
\implies &&\left((2(\D_p+1)\b_{12}^{p\{1,1\},0}+3\b_{12}^{p\{0,(0,1)\},0})M_{-1} + \left(2(2\D_p+1)\b_{12}^{p\{2,0\},0}\right.\right. \cr
&&\left.\left. +3\b_{12}^{p\{(0,1),0\},0}) L_{-1}\right)\right)|\D_{p},0\>=	\frac{(1+\D_{p})}{2}L_{-1}|\D_{p},0\>
\eea
which gives
\be 
 2(1+2\D_{p})\b_{12}^{p\{2,0\},0}+3\b_{12}^{p\{(0,1),0\},0}=\frac{1}{2}(1+\D_{p}),
\label{vir1}
\ee
\be
2(1+\D_{p})\b_{12}^{p\{1,1\},0}+3\b_{12}^{p\{0,(0,1)\},0}=0.
\label{vir2}
\ee
We also have
\bea 
&& L_{2}|2,0\>_p	=	(\D+\D_{p})|0,0\>_p \cr
\implies && \left(6\D_{p}\b_{12}^{p\{2,0\},0}+\b_{12}^{p\{(0,1),0\},0}\left(4\D_{p}+\frac{1}{2}c_L\right)\right)|\D_{p},
0\>	=	(\D+\D_{p})|\D_{p},0\>\cr
\implies &&6\D_{p}\b_{12}^{p\{2,0\},0}+\b_{12}^{p\{(0,1),0\},0}\left(4\D_{p}+\frac{1}{2}c_L\right)=(\D+\D_{p}).
\label{vir3}
\eea
Solving \eqref{vir1} and \eqref{vir3} we have
\be
\b_{12}^{p\{2,0\},0}=\frac{c_L-12\D-4\D_{p}+c_L\D_{p}+8\D_{p}^{2}}{4(c_L-10\D_{p}+2c_L\D_{p}+16\D_{p}^{2})}
\ee 
\be
\b_{12}^{p\{(0,1),0\},0}=\frac{2\D-\D_{p}+4\D\D_{p}+\D_{p}^{2}}{c_L-10\D_{p}+2c_L\D_{p}+16\D_{p}^{2}}.
\ee 
We also have
\be
\b_{12}^{p\{1,1\},1}=-\frac{1}{2}\b_{12}^{p\{2,0\},0}=-\frac{1}{2}\frac{c_L-12\D-4\D_{p}+c_L\D_{p}+8\D_{p}^{2}}{4(c_L-10\D_{p}+2c_L\D_{p}+16\D_{p}^{2})},
\ee 
\be
\b_{12}^{p\{0,(0,1)\},1}=-\frac{1}{2}\b_{12}^{p\{(0,1),0\},0}=-\frac{1}{2}\frac{2\D-\D_{p}+4\D\D_{p}+\D_{p}^{2}}{c_L-10\D_{p}+2c_L\D_{p}+16\D_{p}^{2}},
\ee 
\be
\b_{12}^{p\{0,2\},2}=-\frac{1}{2}\b_{12}^{p\{1,1\},1}=\frac{1}{4}\frac{c_L-12\D-4\D_{p}+c_L\D_{p}+8\D_{p}^{2}}{4(c_L-10\D_{p}+2c_L\D_{p}+16\D_{p}^{2})},
\ee 
and the coefficients $\b_{12}^{p\{0,2\},1},\,\b_{12}^{p\{1,1\},0},\,\b_{12}^{p\{0,2\},0},\,\b_{12}^{p\{0,(0,1)\},0}$ are undetermined.  
All the coefficients other than $\b_{12}^{p\{2,0\},0}$ and $\b_{12}^{p\{(0,1),0\},0}$ are coefficients arising from null states and their descendants. In the chiral truncation of the BMS algebra, all of these should just be ignored. It is true that in an ideal situation, we should have found that either these were zero or undetermined in the limit. We don't understand this aspect of our results completely.

\end{document}